\documentclass[usenatbib,dvipsnames]{mnras}
\usepackage[utf8x]{inputenc}
\usepackage[]{natbib}
\usepackage[T1]{fontenc}
\usepackage{aecompl}
\usepackage{rotating}
\usepackage{stmaryrd}
\usepackage{aas_macros}
\usepackage{times}
\usepackage{graphicx,subfig}
\usepackage{amssymb}
\usepackage{amsmath}
\usepackage{amsfonts}
\usepackage{placeins}
\usepackage{todonotes}
\usepackage{paralist}
\usepackage{multirow}
\usepackage{hyperref}
\usepackage{xcolor}
\usepackage{soul}
\usepackage{units}
\usepackage[normalem]{ulem}

\def\gsim { \lower .75ex \hbox{$\sim$} \llap{\raise .27ex \hbox{$>$}}}
\def\lsim { \lower .75ex \hbox{$\sim$} \llap{\raise .27ex \hbox{$<$}}}

\newcommand{\lya}{{\hbox{\rm Lyman-$\alpha$}}}

\begin{document}

\graphicspath{{fig/}} 

\title[Lyman-$\alpha$ forest and WDM]{The Lyman-$\alpha$ forest as a
  diagnostic of the nature of the dark matter}

\author[A.~Garzilli et al.]  {Antonella Garzilli$^{1}$\thanks{E-mail:
    garzilli@nbi.ku.dk}, 
Andrii Magalich$^{2}$, Tom Theuns$^3$, Carlos S.~Frenk$^3$, \newauthor
Christoph Weniger$^{4}$, Oleg Ruchayskiy$^{1}$ \& Alexey Boyarsky$^{2}$
\\
  $^1$ Discovery Center, Niels Bohr Institute, Copenhagen University, Blegdamsvej 17, DK-2100 Copenhagen, Denmark\\
  $^2$ Lorentz Institute, Leiden University, Niels Bohrweg 2,
  Leiden, NL-2333 CA, The Netherlands\\
$^3$ Institute for Computational Cosmology, Department of Physics, University of Durham, South Road, Durham, DH1 3LE, UK\\
   $^4$ GRAPPA, Institute of Physics, University of Amsterdam,
  Science Park 904, 1098XH Amsterdam, The Netherlands\\
} 
\date{Accepted --. Received --; in original form --}
\maketitle

\begin{abstract}
The observed \lya\ flux power spectrum (FPS) is suppressed on scales
below $\sim~ 30~{\rm km~s}^{-1}$. This cutoff could be due to the high
temperature, $T_0$, and pressure, $p_0$, of the absorbing gas or,
alternatively, it could reflect the free streaming of dark matter
particles in the early universe. We perform a set of very high
resolution cosmological hydrodynamic simulations in which we vary
$T_0$, $p_0$ and the amplitude of the dark matter free streaming, and
compare the FPS of mock spectra to the data. We show that the location
of the dark matter free-streaming cutoff scales differently with
redshift than the cutoff produced by thermal effects and is more
pronounced at higher redshift. We, therefore, focus on a comparison to
the observed FPS at $z>5$. We demonstrate that the FPS cutoff can be
fit assuming cold dark matter, but it can be equally well fit assuming
that the dark matter consists of $\sim 7$~keV sterile neutrinos in
which case the cutoff is due primarily to the dark matter free
streaming.
\end{abstract}


\section{Introduction}
\label{sec:introduction}
The $\Lambda$CDM cosmogony provides an excellent description of the
statistical properties of the cosmic microwave background (CMB),
relating the temperature fluctuations detected in the CMB to the
density fluctuations in the distribution of galaxies (see {\em e.g.}
\citet{Planck18} for a recent description). Non-baryonic \lq dark
matter\rq\ (DM) is a crucial ingredient of the model, reconciling the
low amplitude of the temperature fluctuations in the CMB with the high
amplitude of fluctuations detected in the total matter density
inferred from the clustering of galaxies.

The detailed properties of the DM particle have little impact on the
success of the $\Lambda$CDM model on large scales, but observations on
small scales could potentially distinguish between rival particle
physics models of the nature of the particle. Depending on how the
dark matter particle is produced in the early universe, intrinsic --
as opposed to gravitationally induced -- DM velocities may strongly
suppress the amplitude of matter fluctuations on scales below a
characteristic \emph{free streaming} length, {$\lambda_{\rm DM}$} (see
{\em e.g.} the discussion by \citet{Boyarsky09a}). DM particles for
which {$\lambda_{\rm DM}$} is of the order of a co-moving megaparsec
(cMpc, where the c in cMpc stresses the fact that the length scale is
a co-moving rather than proper quantity and that is measured in Mpc
rather than in Mpc/$h$, that has been the customary unit) are called
\textit{warm dark matter} (WDM). Sometimes WDM refers to the specific
case where the DM is produced in thermal equilibrium, in which case
there is a one-to-one relation between {$\lambda_{\rm DM}$} and the DM
particle mass, $m_{\rm DM}$ (the smaller $m_{\rm DM}$, the larger
{$\lambda_{\rm DM}$}). Both {$\lambda_{\rm DM}$} and $m_{\rm DM}$ can
then be used to quantify the \lq warmness\rq\ of the DM.

The effects of free-streaming on structure formation may be detectable
if {$\lambda_{\rm DM}$} is large enough. Particle free-streaming
introduces a maximum phase-space density of fermionic DM which could
potentially cause dark matter halos to have a central density \lq
core\rq\ \citep{Tremaine79,Maccio12,Shao13}. The smallness of such a
core \citep{Shao13}, and the potential for baryonic processes
associated with star formation and gas cooling to affect the central
density profile (see {\em e.g.}
\cite{Navarro96,Governato10,Pontzen12}), render this route to
determining {$\lambda_{\rm DM}$} challenging \citep{Oman15}. A large
value of {$\lambda_{\rm DM}$} will also dramatically reduce the
abundance of low-mass DM halos (see {\em e.g.}
\cite{Schneider13,Angulo13}) and consequently also of the low-mass
(\lq dwarf\rq) galaxies they host. The abundance of Milky Way
satellites, for example, therefore provides interesting limits on
{$\lambda_{\rm DM}$} \citep{Lovell16,Lovell17}. However the impact of
relatively poorly understood baryonic physics may ultimately limit the
constraining power of both methods. Methods that are largely free from
such uncertainties are therefore more promising; these include
gravitational lensing by low-mass halos \citep{Li16}, and the creation
of gaps in stellar streams by the tidal effects of a passing dark
matter subhalo \citep{Erkal16}. The method for constraining
{$\lambda_{\rm DM}$} that we consider in this paper is based on the
small-scale cut-off in the flux power spectrum of the \lya\ forest.

Residual neutral hydrogen gas in the intergalactic medium (IGM)
produces a series of absorption lines in the spectrum of a background
source such as a quasar, through scattering in the $n=1\rightarrow 2$
\lya\ transition (see {\em e.g.} the review by \citet{Meiksin09}). The
set of lines for which the columndensity of the intervening absorber
is low, $N_\ion{H}{1}\le 10^{16}{\rm cm}^{-2}$, is called the
\lya\ forest. The transmission $F$, {\em i.e.} the fraction of light
of the background source that is absorbed, is often written in terms
of the optical depth $\tau$, as $F=\exp(-\tau)$; we will refer to this
quantity that is independent of the quasar spectrum and only depends
on the intervening distribution of neutral gas, as the
flux\footnote{Let ${\cal F}$ be the observed quasar flux, and ${\cal
    C}$ what would be the observed flux in the absence of absorption,
  then $F\equiv {\cal F}/{\cal C}$ is the transmission. This quantity
  is commonly but somewhat inaccurately referred to as the \lq
  flux\rq, we will do so as well. Since ${\cal C}$ is not directly
  observable, neither is $F$. Estimating $F$ from ${\cal F}$ is called
  \lq continuum fitting\rq.}. The observed power spectrum of $F$
exhibits a cut-off on scales below $\lambda_F\approx 30\,{\rm
  km~s}^{-1}$ {at high redshift}, and currently provides the most
stringent constraints on {$\lambda_{\rm DM}$}
\citep{Hansen01,Viel05,Viel06,Seljak06,Boyarsky09a,Viel13,Baur16,baur2017}.
The reason that the \lya\ forest provides such tight constraints on
{$\lambda_{\rm DM}$} is that the neutral gas follows the underlying
dark matter relatively well, because the absorption occurs in regions
close to the cosmological mean density, particularly at higher
redshifts $z\ge 5$. Nevertheless there are complicating factors, which
include:
\begin{itemize}
\item[({\em i})] the density is probed along a single sightline; the measured one dimensional (1D) power spectrum is an integral of the 3D underlying matter power spectrum (as discussed in details in Appendix~\ref{app:shape});
\item[({\em ii})] the flux is related to the density by a non-linear transformation \citep{miralda1993};
\item[({\em iii})] absorption lines are Doppler broadened;
\item[({\em iv})] the gas distribution is smoothed compared to the dark matter due to its thermal pressure \citep{Gnedin98}.
\end{itemize}
As a consequence, {$\lambda_{\rm DM}\ne \lambda_F$}, and numerical
simulations that try to account for all these effects are used to
infer {$\lambda_{\rm DM}$} by calculating mock absorption spectra, and
comparing $\lambda_F$ from the simulations to the observed
value. However, the temperature of the gas, and hence the level of
Doppler broadening, $\lambda_b$, that needs to be applied, is not
accurately known (see {\em e.g.} \citet{Garzilli15, Rorai18}),
especially at higher redshifts, $z\gtrapprox 5$, where the density
field is more linear which makes it easier to simulate the IGM more
accurately. The smoothing due to gas pressure \citep{Theuns00} can be
described in linear theory \citep{Gnedin98} and the smoothing scale,
$\lambda_p$, depends on the thermal {\em history} of the gas; that
history is not well constrained.

The temperature of the gas is thought to result from a balance between
photoionisation heating and adiabatic cooling \citep{Hui:1997dp,
  Theuns98}. This results in a tight power-law relation between gas
temperature and density, the \emph{temperature-density} (or $T-\rho$)
\emph{relation}: {
    \begin{equation}
\label{eq:TDR}
T = T_0 \left(\frac{\rho}{\bar\rho}\right)^{\gamma-1},
\end{equation}
where $\bar\rho$ denotes the mean density.}

In terms of the smoothing scales discussed above, the value of
$\lambda_b$ at a given redshift $z=z_1$ depends on the parameters of
this temperature-density relation at $z=z_1$, but the value of
$\lambda_p$ depends on the history, $T_0(z)$ and $\gamma(z)$ for $z\ge
z_1$.

Inferring {$\lambda_{\rm DM}$} from $\lambda_F$ then requires running
a number of simulations with different histories, $T_0(z)$ and
$\gamma(z)$, and finding a set of simulations that yield the best
agreement between the simulated and observed value of $\lambda_F$,
while being consistent with observational constraints on the evolution
of $T_0(z)$ and $\gamma(z)$. However constraints on the latter are not
very tight (see {\em e.g.} \citet{madau2017} for a recent discussion
on the nature and evolution of the sources of ionising
radiation). Since we expect that, approximately, $\lambda_F^2\approx
\lambda_b^2+\lambda_p^2+{\lambda_{\rm DM}^2}$ (as would be the case in
the linear regime {\citep{huignedinzhang1997}}), we apply the
following strategy in this paper: {\em we perform simulations with
  $\lambda_p\approx 0$, and examine how well simulations with a given
  $(\lambda_b,{ \lambda_{\rm DM}})$ reproduce the observed value of
  $\lambda_F$.}  We believe that this method yields a robust upper
limit on {$\lambda_{\rm DM}$}. Furthermore, we demonstrate with
simulations that do include photoheating at a level that is consistent
with current constraints, that WDM models with our inferred limit on
$\lambda_{\rm DM}$ are indeed consistent with all current data.

We also specialise to a particular DM candidate~-- \emph{sterile
  neutrinos, resonantly produced in the presence of a lepton
  asymmetry} \citep{Shi:1998km,Laine:2008pg}.  If such a sterile
neutrino (SN in what follows) is sufficiently light (masses of the
order $m_{\rm DM}c^2\approx {\rm keV }$), the 3D linear matter power
spectrum exhibits a cutoff below a scale {$\lambda_{\rm DM}$} that is
a function of two parameters: the mass of the particle, $m_{\rm
  DM}\equiv m_{\rm SN}$, and the primordial lepton asymmetry parameter
that governs its resonant production, $L_6$ (see
e.g. \citet{Laine:2008pg, Boyarsky:2008mt, Lovell16}); see {\em e.g.}
\citet{Boyarsky:2018tvu} for a review on keV sterile neutrinos as a DM
candidate.

\section{The observed flux power spectrum}
\label{sec:data}
In this paper we compare our simulation results to the same flux power
spectrum (FPS) computed from a set of $z\gtrsim 4.5$ quasar spectra
previously analysed by
\cite{Viel13,Garzilli:2015iwa,irsic2017,Irsic:2017yje}, and
\cite{Murgia:2018now}. These data are based on 25 high-resolution
quasar spectra with emission redshifts in the range $4.48\leq z_{{\rm
    QSO}} \leq 6.42$ obtained with the {\sc hires} spectrograph on
{\sc keck}, and the Magellan Inamory Kyocera Echelle ({\sc mike})
spectrograph on the Magellan Clay telescope. We do not analyse the
original spectra -- they are not yet publicly available -- but simply
compare to the published FPS.  We note that for $z= 5.0$ MIKE dataset
contains 4 QSOs with the emission redshifts $z >
4.8$~\citep{becker2011,calverley2011}, while the HIRES dataset
consists of 16 QSOs \citep{becker2007,becker2011,calverley2011}.  At
this redshift the interval $\Delta z = 0.4$ used for binning in
\citet{Viel13} corresponds to $\sim \unit[140]{Mpc}/h$.  Taking into
account quasar proximity zones these quasar spectra cover $\sim
\unit[240]{Mpc}/h$ (MIKE) and $\unit[1230]{Mpc}/h$ (HIRES) at $z=5$
and $\sim\unit[810]{Mpc}/h$ for {\sc hires} at $z=5.4$.  From this we
can already anticipate that the sample variance errors will be quite
large for both datasets.  We will use this information in
Section~\ref{sec:thermal_history} below when estimating errors due to
this finite sampling.

The {\sc hires} and {\sc mike} spectra have a spectral resolution of
6.7 and 13.7 ${\rm km~s}^{-1}$ full width at half maximum (FWHM), and
pixel size of 2.1 and $5.0\,{\rm km~s}^{-1}$, respectively. The median
signal-to-noise ratios at the continuum level are in the range 10--20
per pixel~\citep{Viel13}. We generate mock FPS with similar
properties, as described below. The finite spectral resolution
introduces another cut-off scale in the FPS, $\lambda_s\sim {\rm
  FWHM}$.

The ionisation level of the IGM is quantified by the \emph{effective
  optical depth}, $\tau_{\rm eff}\equiv -\ln\langle F \rangle$, where
$\langle F \rangle$ is the observed mean transmission, averaged over
all line-of-sights. \citet{Viel13} report values of $\tau_{\rm
  eff}(z=5.0)=1.924$ and $\tau_{\rm eff}(z=5.4)=2.64$, without quoting
associated uncertainties which can be quite large, stemming from the
systematic errors in continuum fitting and statistical errors due to
sample variance.  {We provide our own estimates of the statistical
  errors due to sample variance on $\langle F \rangle$ in
  Appendix~\ref{app:meanfluxerror}.}  For details on the properties of
the dataset, the associated noise level, and the way the FPS and its
covariance matrix were estimated, we refer the reader to
\citet{Viel13}.

\section{Flux power spectrum}
\label{sec:simulated_FPS}
As outlined above, in the present paper we compare the mock FPS
computed from simulations to the observed FPS presented by
\cite{Viel13}. Traditionally the FPS is computed in \lq velocity
space\rq. Integrating the Doppler shift relation between wavelength
and velocity, $dv/c=d\lambda/\lambda$, the redshift or wavelength
along a line-of sight to a quasar can written in terms of a \lq
Hubble\rq\ velocity $v$ as \begin{equation}
\label{eq:vl}
  v = c~\ln\left({\frac{\lambda}{\lambda_0(1+z)}}\right)=\frac{ H(z)}{1+z} y\,,
\end{equation}
where $\lambda_0=1215.67\,$\AA\, is the laboratory wavelength of the
\lya\ transition, and $z$ is a constant reference redshift. The
zero-point of $v$ is defined by $z$ and is arbitrary. In data, $z$ is
often chosen to be the mean redshift of the data or the quasar's
emission redshift, in simulations we take it to be the redshift of the
snapshot. In this equation, $H(z)$ is the Hubble constant at redshift
$z$, and the right-hand side also defines a co-moving position $y$
along the spectrum.

The input to the FPS (either observed or obtained from simulations) is
then flux as function of velocity, {\em i.e.} $F(v)$, over some
velocity interval $V$ (in the data set this interval is chosen so that
one avoids the Lyman-$\beta$ forest, the quasar near zone, and
potentially some strong absorbers; in the simulations it is set by the
linear extent of the simulated volume).

Given $F$ and its mean, $\langle F\rangle$, we calculate the \lq normalised flux\rq\,
\begin{eqnarray}
	\delta_F \equiv \frac{F-\langle F \rangle}{\langle F \rangle}\,.
\end{eqnarray} 
The FPS is written in terms of the dimensionless variance $\Delta_F^2(k)$ (strictly speaking a variance in $\delta_F$ per dex in $k$), defined by
\begin{align}
\Delta_F^2(k) &= \frac{1}{\pi} k P_F(k)\\
P_F(k) &= V \left\langle |\tilde{\delta}_F(k)|^2 \right\rangle \\
\tilde{\delta}_F(k) &= \frac{1}{V} \int_0^V dv\, e^{-ikv} \delta_F(v)\,.
\end{align}
Here, $\langle\cdot\rangle$ denotes the ensemble average, and $k =
2\pi / v$ is the Fourier \lq frequency\rq\ corresponding to $v$ and
has dimensions of (s/km). To find the conversion to a wave-vector in
inverse co-moving Mpc, $k_x$, recall that the Hubble law of
Eq.~(\ref{eq:vl}) states that $\Delta v=H(z)\Delta y/(1+z)$. Then,
since $k_y\,y=k_v\,v$, where $k_v\equiv k$, we find that
\begin{equation}
  k_y = k_v {H(z)\over 1+z}\,.
\end{equation}

The aim of the analysis is to identify the smoothing lengths defined
in the \nameref{sec:introduction}, {\em i.e.} $\lambda_b$ -- the
Doppler broadening, $\lambda_p$ -- the pressure smoothing, and
$\lambda_{\rm DM}$ -- the dark matter free-streaming length, as a
cut-off in the FPS. Suppose that $\Delta_F^2(k)$ declines rapidly
above a characteristic value of $k$, say $k_{\rm max}$. How is $k_{\rm
  max}$ related to the smoothing length $\lambda$?

The simplest case is that of Doppler broadening. Consider a sharp
feature in $F(v)$, smoothed by Doppler broadening due to gas being at
temperature $T$. The width of the smoothed feature in velocity space
will be of order $\Delta v_b=\left(2k_{\rm B}T/ m_{\rm
  H}\right)^{1/2}$ (where $k_{\rm B}$ is Boltzmann's constant and
$m_{\rm H}$ the proton mass). In terms of the Fourier transform of
$F(v)$, this will correspond to a feature at the proper
wavenumber\footnote{This is the case for Gaussian smoothing in the
  linear regime, with the factor 2 arising from the fact that the
  power spectrum is the square of the Fourier transform.}

\begin{equation}
k_{{\rm max},b}={\sqrt{2}\over\Delta v_b}=0.11~\left({T\over 10^4\,{\rm K}}\right)^{-1/2}~({\rm km~s^{-1}})^{-1}\,,
\label{eq:kmaxb}
\end{equation}
which is independent of $z$, provided that $T$ is constant.

How about pressure smoothing? The extent of the smoothing is
approximately of order of the Jeans length \citep{schaye2001}, which
in proper units is
\begin{equation}
	\lambda_J = \sqrt{c_s^2\pi\over {\rm G}\rho}\;.
\end{equation}
Here, $\rho$ is the total mass density (dark matter plus gas) of the
absorber and $c_s$ the sound speed. The corresponding velocity
broadening is then $\Delta v_p=H(z)\lambda_J/(2\pi)$
\citep{Garzilli15}. At high enough redshift, the Hubble parameter
scales like $\propto (1+z)^{3/2}$, and the density dependence of
$\lambda_J$ also scales like $\rho^{-1/2}\propto (1+z)^{3/2}$, making
$\Delta v_p$ also independent of redshift\footnote{We note that this
  no longer true at low redshift, where $\Delta v_b$ and $\Delta v_p$
  scale differently with $z$.}. The corresponding value of $k_{\rm
  max}$ is
\begin{equation}
k_{{\rm max},p}={\sqrt{2}\over \Delta v_p}=0.0760~\left({T\over 10^4{~\rm K}}\right)^{-1/2}\left({\rm km~s^{-1}}\right)^{-1}\,.
\label{eq:kmaxp}
\end{equation}

The width of a feature due to dark matter free-streaming, {
  $\lambda_{\rm DM}$}, is imprinted in the linear transfer function,
and is therefore constant in co-moving (as opposed to proper)
coordinates. The velocity extent of such a feature is therefore
$\Delta v_\lambda=H(z)\lambda_{\rm DM}/(1+z)\propto (1+z)^{1/2}$ at
high-enough $z$, and in the FPS scales like $k_{{\rm max, DM}}\propto
\Delta v_\lambda^{-1}\propto (1+z)^{-1/2}$ and hence is {\em not}
independent of $z$. We can write its value as
\begin{multline}
k_{{\rm max},{\rm DM}}={1+z\over H(z)}\,{1\over \lambda_{\rm DM}}\\
=0.007\,\left({\lambda_{\rm DM}\over h^{-1}{\rm
    cMpc}}\right)^{-1}\left({6\over 1+z}\right)^{1/2}\left({\rm
  km~s}^{-1}\right)^{-1}\; .
\end{multline}
The free-streaming scale $\lambda_{\rm DM}$ can be estimated as a
position of the maximum of the linear matter power spectrum, see
Fig.~\ref{fig:mps}.  For a particular case of 7~keV sterile neutrino
that we will investigate in this work, this scale can be found
e.g.\ in \citet{Lovell16} as a function of lepton asymmetry.  For the
model with lepton asymmetry parameter $L_6 = 12$ \citep[see][for the
  definition of $L_6$]{Boyarsky09a} the resulting scale is
$\lambda_{\rm DM} \sim \unit[0.07]{Mpc}/h$ which corresponds to
$k_{\rm DM,max} \approx \unit[0.1]{sec/km}$ at $z=5$.

Finally, the finite resolution of the spectrograph imprints a feature
that is constant in velocity space since the spectral resolution has a
given value of ${\cal R}\equiv \Delta\lambda/\lambda={c/\Delta
  v_s}$. The feature occurs at the redshift independent wavenumber
\begin{equation}
k_{{\rm max},s}={\sqrt{2} \over \Delta v_s}=0.21\,\left({6.6\,{\rm km~s}^{-1}\over \Delta v_s}\right)^{-1} \,\rm \left(km\,s^{-1}\right)^{-1}\,.
\end{equation}

The conclusion of this is that the effects of free-streaming, compared
to those of thermal broadening, pressure smoothing or finite spectral
resolution, scale differently with $z$.  The redshift dependence is
sufficiently weak so to make little difference between $z=5.4$ and
$z=5$, but the difference does become important comparing the FPS at
$z=3$ versus $z=5$. The numerical values also suggest that
free-streaming, Doppler and pressure broadening set-in at very similar
values of $k$, and that the finite spectral resolution of {\sc keck}
is unlikely to compromise the measurements.

When simulating the above effects using a hydrodynamical simulation,
yet another scale enters: the Nyquist frequency, set by the mean
interparticle spacing. For a simulation with $N^3$ particles in a
cubic volume with linear extent $L$, the corresponding scale is
$\lambda_{\rm sim}=L/N^{1/3}$, and is constant in co-moving units. The
corresponding $k_{\rm max}$ is of order
\begin{equation}
k_{{\rm max, sim}}={(1+z)\over H(z)}\,{N^{1/3}\over L}\approx 0.27~({\rm km~s}^{-1})^{-1}\,,
\end{equation}
where the numerical value is for $z=5$, $L=20\,h^{-1}{\rm Mpc}$ and
$N=512^3$, suggesting that the numerical resolution needs to be at
least this good in order not to compromise the location of any cut-off
in mock spectra. We discuss our numerical simulations next.

\section{Simulated flux power spectra}
\label{sec:modeling}
\subsection{Strategy}
Hydrodynamical cosmological simulations usually expose the gas in the
IGM to a uniform (homogeneous and isotropic) but evolving ionising
background that mimics the combined emissivity of radiation from
galaxies and quasars \cite[see e.g.][]{haardt1996}. As a result, the
mean neutral fraction is very low: $x\equiv n_\ion{H}{I}/n_{\rm H}\ll
1$. Without such an ultraviolet background (UVB), the effective
optical depth would be much higher than observed \citep{gunn1965}.

Assuming that the UVB is uniform may be a good approximation long
after reionisation, when fluctuations around the mean photoionisation
rate, $\Gamma_\ion{H}{I}$, are small \citep{Croft04,McDonald05}.
However, this may no longer be the case closer to reionisation when
the UVB may be much more patchy \cite[e.g.][]{Becker18, Bosman18}. The
current best-estimate for the redshift of reionisation is $z_{\rm
  reion}=7.82\pm 0.71$, with a reionisation history consistent with a
relatively rapid transition from mostly neutral to mostly ionised, and
suggesting the presence of regions that were reionised as late at
$z\sim 6.5$ {\citep{Planck2016}}. These inferences obtained from the
CMB are also consistent with hints of extended parts of the IGM being
significantly neutral, $x\sim 0.1-0.5$, in the spectra of $z\gtrapprox
7$ quasars \citep{Mortlock11, Davies18}. Such late reionisation, and
the patchiness associated with it, make it much harder to perform
realistic simulations of the IGM that yield robust constraints on
$\lambda_{\rm DM}$. In fact, the impact of large fluctuations in
$\Gamma_\ion{H}{I}$ is not just restricted to inducing fluctuations in
$x$, the neutral fraction, because the UVB also heats gas.

The temperature $T$ of a photoionised IGM depends on the density and
on the spectral shape of the ionising radiation \citep{Miralda94,
  Abel99}. Unlike the more familiar case of galactic \ion{H}{II}
regions, $T$ is not set by a balance between photoheating and
radiative cooling, but by the mostly impulsive heating during
reionisation and the adiabatic expansion of the
Universe. Nevertheless, the temperature $T_0$ in the
temperature-density relation of Eq.~(\eqref{eq:TDR}) is expected to be
of the order $T_0\sim 10^4$~K with $\gamma\approx 1$ close to
reionisation. Once heated, pressure will smooth the gas distribution
relative to the underlying dark matter introducing the filtering scale
$\lambda_p$ discussed previously, below which the amplitude of the
density power spectrum is strongly suppressed. The patchiness of
reionisation will therefore introduce large-scale fluctuations in the
neutral fraction $x$, but also in the value of $\lambda_p$, as well as
in that of the Doppler-broadening $\lambda_b$.

Although it is possible to carry out approximately self-consistent
simulation of the IGM during reionisation (e.g. \cite{Pawlik17}), such
calculations are still relatively computationally demanding. We
therefore use the following strategy in this paper: we perform some of
the simulations {\em without} imposing a UVB, meaning that effectively
$\lambda_p=0$.  We then apply an \lq effective\rq\ UVB in
post-processing, by imposing a given temperature-density relation of
the form given by Eq.~\eqref{eq:TDR} and scaling the neutral fraction
$x$ to obtain the observed effective optical depth (as described in
more detail below). We stress therefore that many of our runs are not
realistic, nor are they intended to be. Quite the opposite, we work in
an idealised scenario that allows us to vary individually every
relevant effect separately. In addition to these runs, we also carry
out simulation that do impose a UVB on the evolving IGM - we use these
to demonstrate that our limits on $\lambda_{\rm DM}$ are also valid in
this more realistic scenario.

\subsection{Numerical simulations}
\label{sec:simulations}
\begin{figure}
\includegraphics[width=0.45\textwidth]{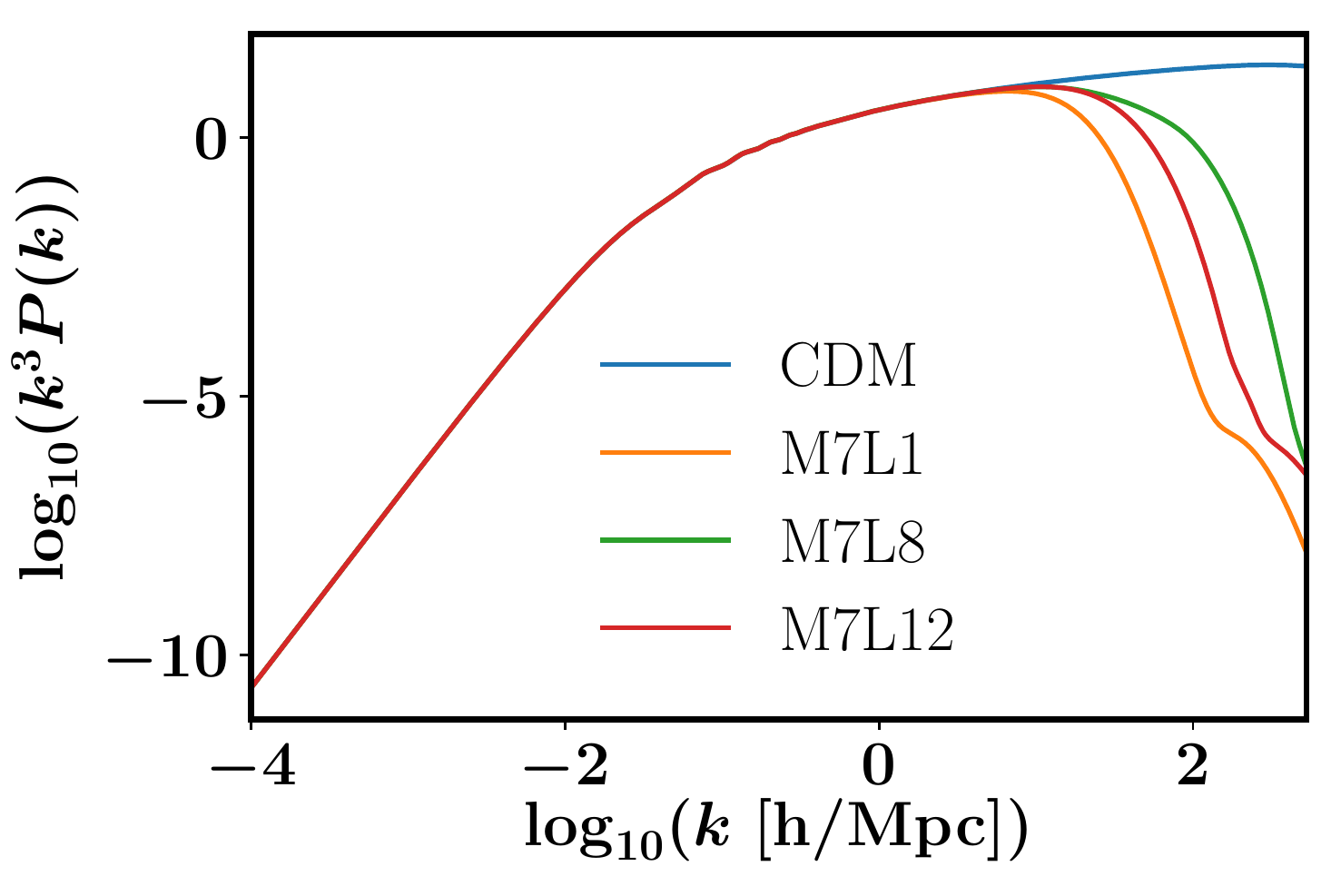}
\caption{Linear dimensionless matter power spectra generated by {\sc
    CAMB} for CDM ({\em blue line}) and for the sterile neutrino model
  with particle mass $m_{SN} = 7 \,\textrm{keV}$ with three different
  choices of the lepton asymmetry parameter $L_6$, as indicated in the
  legend ({\em orange, green and red}, for $L_6=1$, 8 and 12,
  respectively).
\label{fig:mps}
}
\end{figure}
\begin{table*}
\centering
\begin{tabular}{l|l|l|||l|c|c|}
  \hline
  Name & $L\, [{\rm Mpc}/h]$ & $N$ & Dark matter & {UVB} & Cosmology\\
  \hline
  CDM\_L128N64 & 128 & $64^3$ & \multirow{4}{*}{CDM} & \multirow{4}{*}{no UVB} & \multirow{4}{*}{{Viel}}   \\
  CDM\_L20N512 & 20 & $512^3$ & & & \\
  CDM\_L20N896 & 20 & $896^3$ & & & \\
  CDM\_L20N1024 & 20 & $1024^3$ & & & \\
  \hline
  {M7L1} & \multirow{3}{*}{20} & \multirow{3}{*}{$1024^3$} & $m_{SN} = 7\,\textrm{keV}, L_6 = 1$ & \multirow{3}{*}{no UVB} & \multirow{3}{*}{Viel} \\
  {M7L8} & & & $m_{SN} = 7\,\textrm{keV}, L_6 = 8$ & & \\
  {M7L12} & & & $m_{SN} = 7 \textrm{keV}, L_6 = 12$ & & \\
  \hline
  CDM\_Planck\_Late & \multirow{3}{*}{20} & \multirow{3}{*}{$1024^3$}& CDM & {\emph{LateR}} & \multirow{3}{*}{Planck} \\
  {CDM\_Planck\_Early} & & & CDM & {\emph{EarlyR}} & \\
  M7L12\_Planck\_Late & & & $m_{SN} = 7\,\textrm{keV}, L_6 = 12$ & \emph{LateR} & \\ 
  \hline
  EAGLE\_REF & 100 $/ h$ & $1504^3$ & CDM & Eagle & Planck \\
  \hline
\end{tabular}
\caption{\label{tab:simulations} Hydrodynamical simulations
  considered in this work together with corresponding
  parameters. All simulations were performed specifically
    for this work, except {\tt EAGLE\_REF} \citep{schaye2015}. 
  Columns contain from left to right: simulation identifier, co-moving
  linear extent of the simulated volume ($L$), number of dark matter
  particles ($N$, there is an equal number of gas particles), type of
  dark matter (CDM or sterile neutrino WDM with the indicated particle
  mass, $m_{\rm SN}$ { -- expressed in natural units --} and lepton
  asymmetry parameter, L$_6$), ultra-violet background imposed during
  the simulation (no UVB indicates no UVB was imposed; \emph{LateR}
  and \emph{EarlyR} refer to the UVBs from the \emph{LateR} and
  \emph{EarlyR} reionization models in
  \protect\citet{onorbe2016}, \emph{Eagle} indicate the
    standard UVB from \citep{haardt2001}), choice of cosmological
  parameters from Table~\protect\ref{tab:cosmology}, and figure where
  the particular simulation is used. The gravitational softening
  length for gas and dark matter is kept constant in co-moving
  coordinates at 1/30$^{\rm th}$ of the initial interparticle
  spacing. All simulations were started from the initial conditions
  generated by the \texttt{2LPTic} \protect\citep{scoccimarro2012}
  with the same \lq glass\rq\-like particle distribution generated by
  GADGET-2 \protect\citep{springel2005}.}
\end{table*}

\begin{table}
\centering
\begin{tabular}{lll}
\hline
Cosmology & \emph{Planck} \citep{Planck2015} & \emph{Viel} \citep{Viel13} \\
\hline
$\Omega_0$       & $0.308 \pm 0.012$  & $0.298$  \\
$\Omega_\Lambda$ & $0.692 \pm 0.012$  & $0.702$  \\
$\Omega_b h^2$   & $ 0.02226 \pm 0.00023$ & $0.022393$ \\
$h$              & $0.6781 \pm 0.0092$ & $0.7$ \\
$n_s$            & $0.9677 \pm 0.0060$ & $0.957$ \\
$\sigma_8$       & $0.8149 \pm 0.0093$ & $0.822$ \\
\hline
\end{tabular}
\caption{\label{tab:cosmology}Cosmological parameters used in our
  simulations. \emph{Planck} cosmology is the conservative choice of
  TT+lowP+lensing from \citet{Planck2015} (errors represent $68\%$
  confidence intervals), while \emph{Viel} cosmology corresponds to
  the bestfit model in \citet{Viel13}.}
\end{table}
In this work, we have considered a suite of
dedicated cosmological hydrodynamical
simulations, and one of the simulations from the Eagle
  simulation suite. Our dedicated simulation suite has been performed
using the simulation code used by \citet{Viel13b}. This code is {a
  modified version of the publicly available {\sc gadget-2} {\sc
    TREEPM/SPH} code described by \cite{springel2005}}; the runs
performed are summarised in Table~\ref{tab:simulations}. The values of
the cosmological parameters used are in Table~\ref{tab:cosmology};
runs labelled \lq Planck\rq\ use parameters taken from
\citet{Planck2015}, those labelled \lq Viel\rq\ use parameters taken
from \citet{Viel13} to allow for a direct comparison with the latter
work.

Initial conditions for the runs were generated using the
\texttt{2LPTic} code described by \cite{scoccimarro2012}, for a
starting redshift of $z=99$ that guarantees all sampled waves are
still in the linear regime.  {The initial linear power spectrum for
  the CDM cosmology was obtained with the linear Boltzmann solver
  CAMB~\citep{Lewis:1999bs}. Sterile neutrino dark matter is also
  modelled as non-interacting massive particles, with the effects of
  free streaming imprinted in the initial transfer function as
  computed with the modified CAMB code described by
  \citet{Boyarsky:2008mt}, using the primordial phase-space
  distribution functions for sterile neutrinos computed
  in~\citet{Laine:2008pg}. Using instead results from the most recent
  computations~\citep{Ghiglieri:2015jua,Venumadhav:2015pla} would not
  change our results.  We neglect the effects of peculiar velocities
  of the WDM particles other than the cut-off they introduce in the
  transfer function. The linear matter power spectra for the different
  models used in this paper are shown in Fig.\ref{fig:mps}.

Simulations in the same boxes use the same set of random numbers, this
allows us to compare \lya\ forest spectra between CDM and WDM directly
(see Fig.~\ref{fig:spectra}).

For simulations that include a UVB, we specify the redshift-dependent
values of the photoionisation and photoheating rates for hydrogen and
helium as input parameters.  The version of {\sc gadget} that we use
solves for the radiative heating and cooling of the photoionised gas,
given these input rates.  Imposing the rates of \citet{onorbe2016}
results in a $T-\rho$ relation that is consistent with that of the
latter authors.  We use the same UVB in the SN cosmology as an
\emph{example} of the reionisation history with a small filtering
scale.

SPH (gas) particles are converted to collisionless \lq
star\rq\ particles when they reach an overdensity $\rho/\bar\rho>1000$
provided their temperature $T<10^5$~K. This \lq quick-\lya\rq\ set-up
reduces run time by avoiding the formation of dense gas clumps with
short dynamical times, that would in reality presumably form stars in
a galaxy. We can do so, because the impact of forming galaxies on the
IGM is thought to be small, particularly at high redshifts and for the
low density gas regions to which our analysis is sensitive
\citep{Theuns02,Viel13b}.

The simulation from the Eagle simulation suite, {\tt EAGLE\_REF}, has
CDM cosmology and UVB as the standard choice from \citep{haardt2001},
further details can be found in \citep{schaye2015}. Its boxsize and
number of particle are respectively $L=100\,{\rm cMpc}$} and to
$N_{\rm part}=1504^3$, and its resolution is smaller by a factor $\sim
5$ respect to the resolution of our highest resolution
simulations. This simulation has been considered for estimating the
covariance matrix of the mean FPS.

\subsection{Calculation of mock spectra}
\label{sec:mock}
\begin{figure*}
  \includegraphics[width=\textwidth]{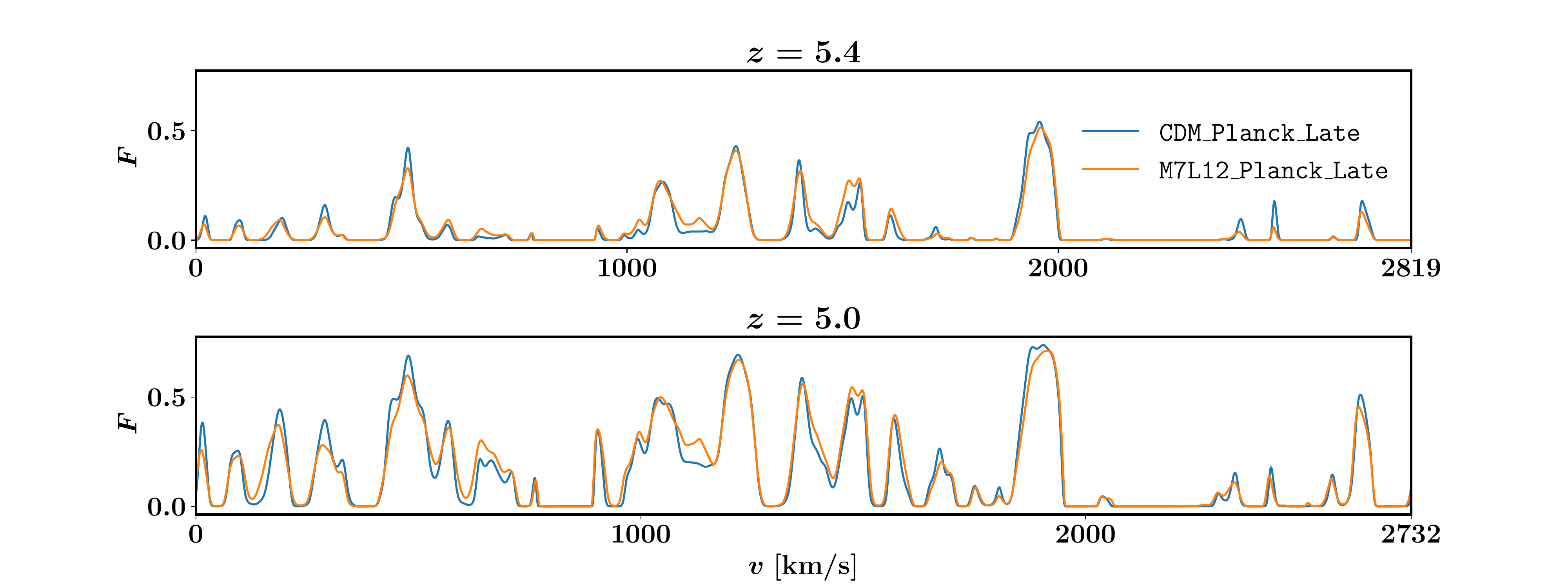}
\caption{Example mock spectra extracted along the same line of sight
  in {\tt CDM\_Planck\_Late} (blue line) and {\tt M7L12\_Planck\_Late}
  (orange line), simulations at redshifts $5.4$ ({\em top panel}) and
  $5.0$ ({\em bottom panel}). The temperature $T_0$ of the gas at the
  mean density at these redshifts is $\sim 7700 \textrm{K}$ for both
  redshifts. Note that a sightline through the full extent of the box
  corresponds to a different velocity extent at different
  redshifts. The evolution of the mean transmission is apparent. The
  CDM and WDM spectra look quite similar, nevertheless on closer
  inspection it is clear that the CDM spectrum has some sharper
  features.}
\label{fig:spectra}
\end{figure*}

\begin{figure*}

\includegraphics[width=0.5\textwidth]{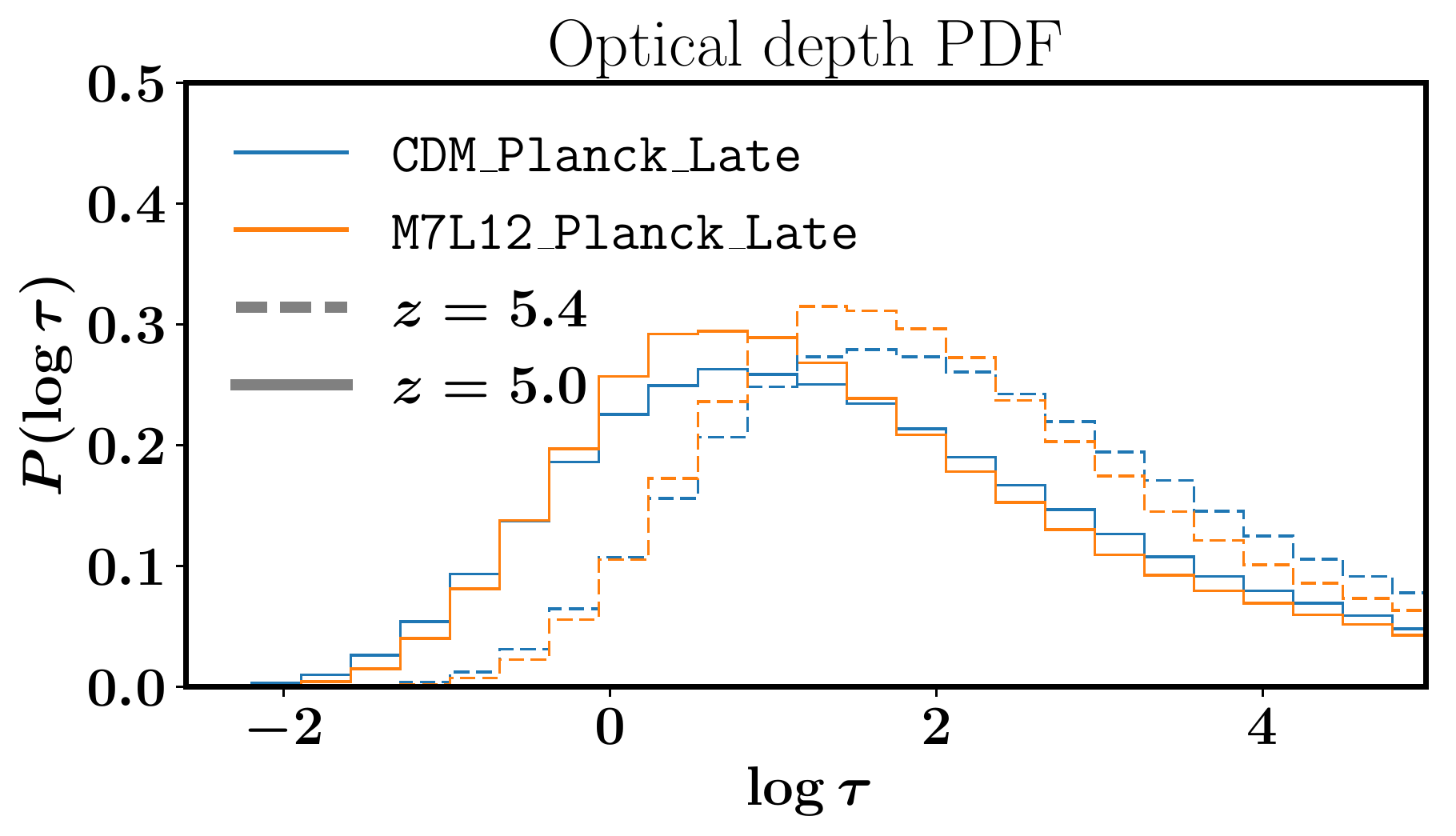}~
\includegraphics[width=0.5\textwidth]{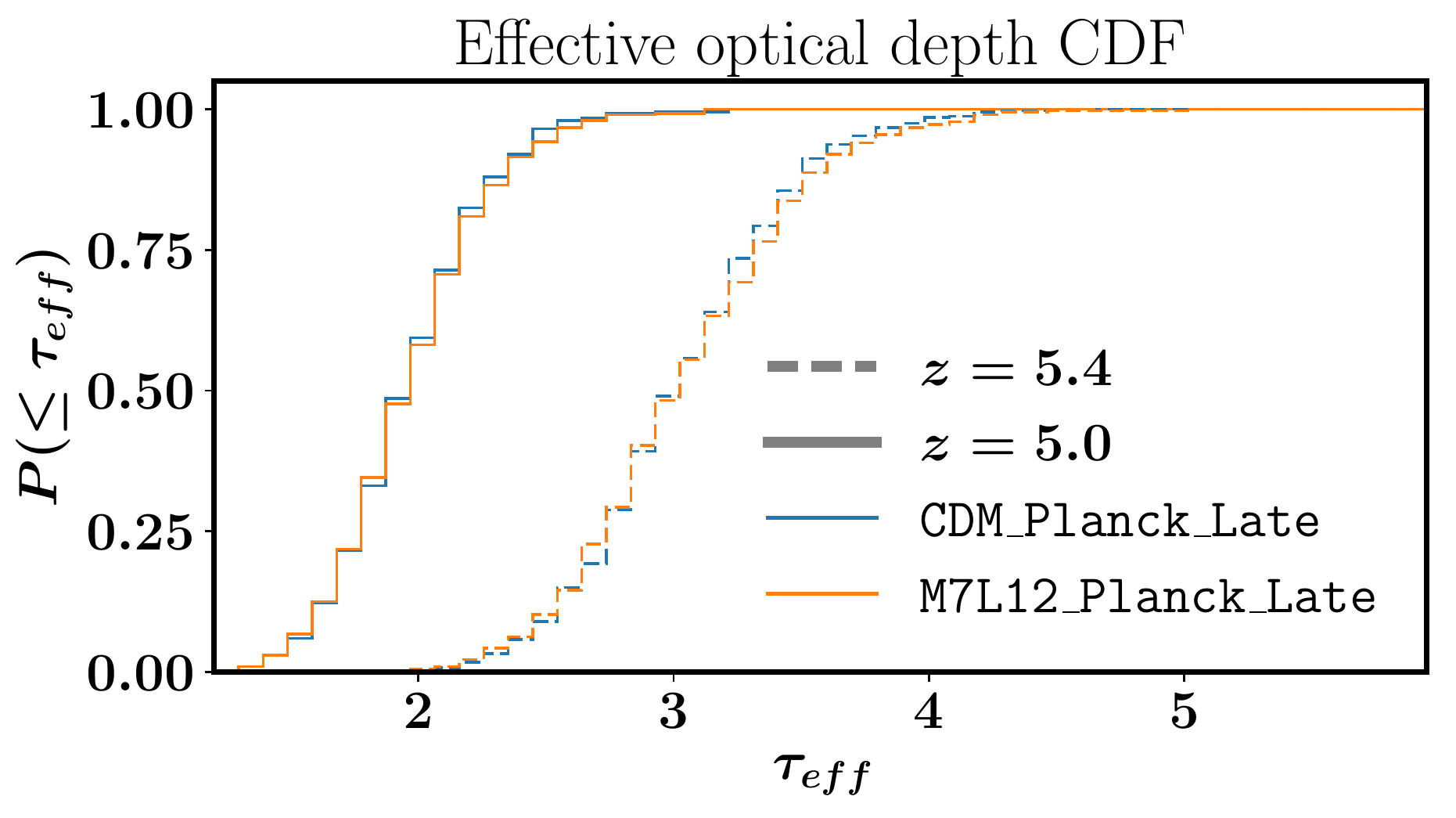}
\caption{{\em Left panel:} probability distribution function of the
  optical depth per pixel. {\em Right panel:} cumulative probability
  distribution of the effective optical depth, $\tau_{\rm eff}$,
  measured in chunks of $50\;\textrm{Mpc~h}^{-1}$. The {\tt
    CDM\_Planck\_Late} model is plotted in {\em blue}, the {\tt
    M7L12\_Planck\_Late} in {\em orange}, redshift $z=5.4$ corresponds
  to {\em dashed lines} and $z=5.0$ to {\em full lines}.}
\label{fig:PDFs}
\end{figure*}

\begin{figure*}
  \centering 
  \includegraphics[width=\columnwidth]{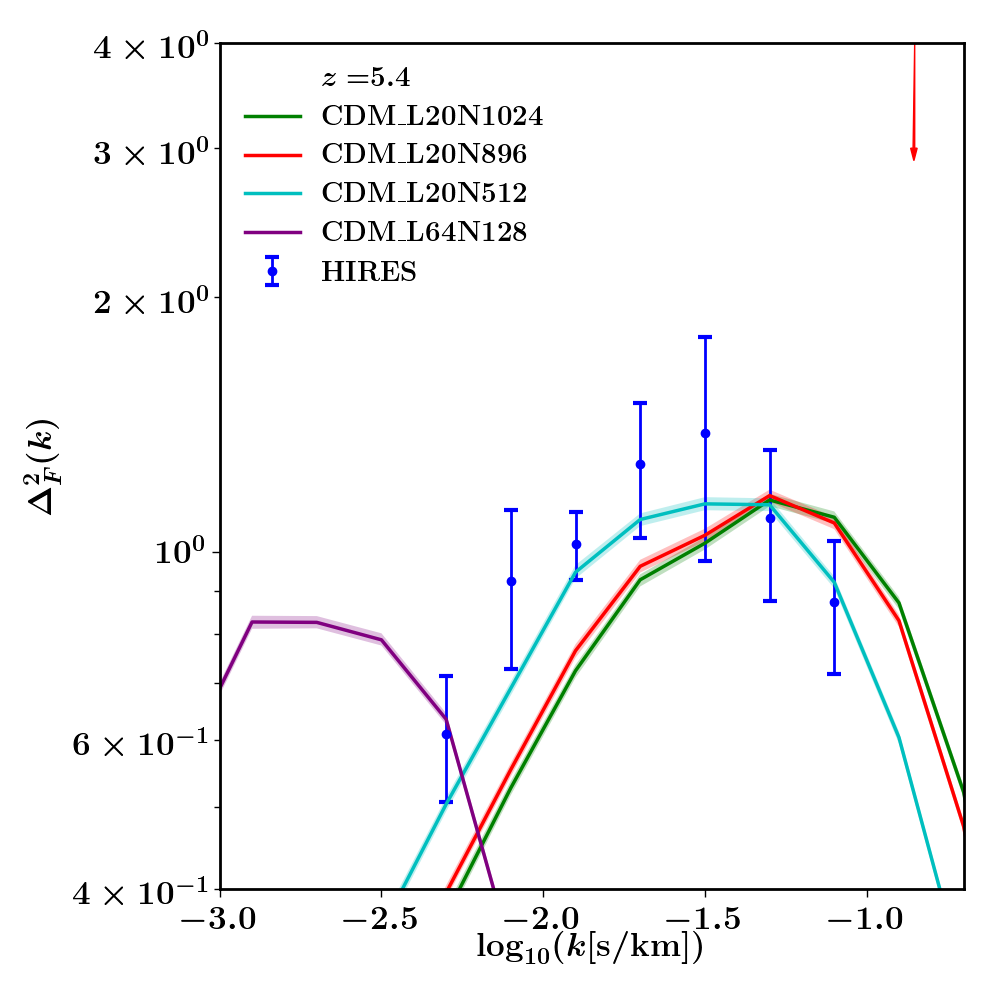} 
  \includegraphics[width=\columnwidth]{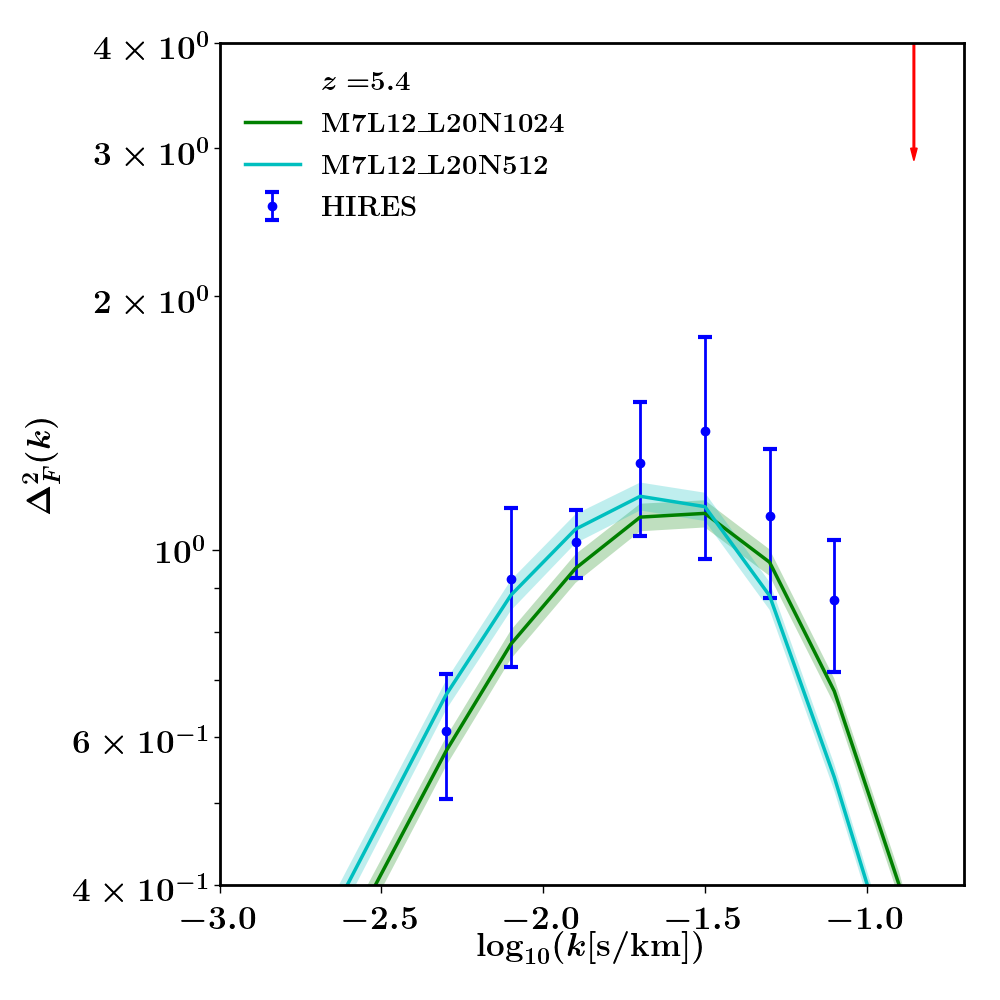}
  \caption{Effect of numerical resolution on the mock FPS for CDM
    ({\em left panel}) and WDM ({\em right panel}) of simulations
    performed without an imposed UVB. Both models are for the imposed
    power-law $T-\rho$ relation of Eq.~\protect\eqref{eq:TDR} with
    $(T_0,\gamma)=(25~{\rm K},1)$, are scaled to the observed value of
    the effective optical depth, $\tau_{\rm eff}=3.0$ for $z=5.4$, and
    mimic the spectral resolution and pixel size of the {\sc hires}
    spectrograph on the {\sc keck} telescope (FWHM=6.7~km~s$^{-1}$,
    pixel size=2.1~km~sec$^{-1}$ (see Section~\ref{sec:data}) but
    without adding noise.  The data points show the error bars as
    reported by \protect\citet{Viel13} that do not take into account
    sample variance (see below).  The different colours correspond to
    different numbers of particles $N$, as per the legend.  The
    observed FPS from \protect\cite{Viel13} (blue) is plotted to
    indicate the range of relevant wave numbers.  There is a numerical
    resolution-dependent cut-off in each simulation.  Increasing the
    number of particles, the position of this cut-off shifts to larger
    $k$ values. In our highest resolution simulations, $N=1024^3$ DM
    and gas particles (green line), the resolution-dependent cutoff is
    outside the range of scales probed by the Lyman-$\alpha$ data, the
    corresponding Nyquist scale $k_{\rm max, sim}$ is outside the
    boundary of the plot.  Therefore, we use such resolution in all
    subsequent simulations.  The red arrow shows the scale associated
    with $k_{\rm max, DM}$.  The figure also demonstrates that the
    simulations considered by \protect\citet{mo_jing_borner1997} ({\em
      purple line}) lacked the necessary resolution to be used in
    \protect\citet{desjacques2004}.  }
\label{fig:check_mo_jing_borner}
\end{figure*}

We compute mock spectra of the simulations using the {\sc specwizard}
code that is based on the method described by \citet{Theuns98}. This
involves computing a mock spectrum along a sight line through the
simulation box along one of the coordinate axis.

For simulations without a UVB ({\tt CDM\_L20N1024}, {\tt M7L12}), we
first impose a temperature-density relation of the form of
Eq.~\eqref{eq:TDR} on all gas particles.  At the high redshifts that
we are considering, the \lya\ transmission is non-negligible only for
sufficiently small overdensities, $\delta \lesssim 1$.  We checked
explicitly that the effect of cooling at the highest densities is
negligible for our analysis.  Therefore, one can safely apply the
temperature-density relation to the whole range of densities
considered, without worrying about it being applicable only in the
range $\delta \lesssim 10$ \citep{Hui:1997dp}.

We use the same post-processing also for simulations which do include
a UVB.  The rationale behind this is the following.  As already
mentioned, we use \citet{onorbe2016} ionisation history only as an
example of the model with small pressure effects, not as a holistic
model.  We then vary the $T_0$ in post-processing (see
Section~\ref{sec:chi2} below) and determine the range of admisssible
temperatures in CDM and WDM cosmologies.  We verify {\it a posteriori}
that the actual temperature predicted by the \emph{LateR} model lies
within the range of admissible temperatures.

Given $T$ and $\rho$ of each particle, we compute the neutral fraction
$x$ using the interpolation tables from \citet{wiersma2009a}, which
assume photoionisation equilibrium,
\begin{equation}
  {dn_\ion{H}{1}\over dt}=-\Gamma_\ion{H}{1} n_\ion{H}{1}-\Gamma_c\,n_e\,n_\ion{H}{1}+\alpha(T)\,n_e\,n_\ion{H}{II}=0\,.
\end{equation}
Here the terms from left to right are photoionisation by the imposed
UVB, collisional ionisation, and recombination (with $\alpha(T)$ the
temperature-dependent case-A recombination coefficient); $n_e$ is the
electron density; the photoionisation rate is that described by
\cite{haardt2001}.

We then interpolate the temperature, density, and peculiar velocity to
the sight line in bins of $\Delta v=1$~km~s$^{-1}$ using the Gaussian
method described by \cite{Altay13}. We verified that this spectral
resolution is high enough to give converged results. We then compute
the optical depth as function of wavelength, $\tau(v)$, thus
accounting for Doppler broadening and the effects of peculiar
velocities.

To allow for a fair comparison to the observed spectra, we convolve
the mock spectra with a Gaussian to mimic the effect of the
line-spread function, and rebin to the observed pixel size with
parameters as described in Section~\ref{sec:data}. The Gaussian white
noise has a uniform relative Standard Deviation of $\sigma=0.066$,
corresponding to a signal to noise ratio of $S/N=15$ per pixel at the
continuum level, following~\cite{Viel13}. Further details on the
application of noise to mock spectra and comparison with previous work
are given in Appendix~\ref{app:noise}. We calculate a set of such
spectra for the snapshot at redshifts $z=5$, and $z=5.4$.

After repeating this procedure for ${\cal N}=10^3$ sight lines, we
compute the mean transmission, $\langle
F\rangle=\langle\exp(-\tau)\rangle$ and scale the optical depth so
that the ensemble of mock spectra reproduces the observed value of
$\langle F\rangle$ discussed in Section~\ref{sec:data}.

We compare spectra along the same sight line for the CDM and the
\texttt{M7L12\_Planck\_Late} models in Fig.~\ref{fig:spectra} (blue
and orange curves, respectively), at redshifts $z=5.4$ (top panel),
and $z=5.0$ (bottom panel); the temperature and thermal history are
the same for both models. The \lya\ spectra look very similar in these
models, although it can be seen that the CDM model has some sharper
features.

The probability distribution function (PDF) of the optical depth is
compared between these two models in Fig.~\ref{fig:PDFs}.

\subsection{Numerical convergence}
\label{sec:convergence}
Before comparing the mock FPS to the observed FPS, we investigate to
what extent the mock FPS is converged, both in terms of resolution and
box size; the latter discussion can be found in
Appendix~\ref{app:box_convergence}. The gas temperature in our
simulations that were performed without an imposed UVB is very low,
and the gas distribution itself is not numerically converged at any of
our resolutions. The effect of that on the FPS is shown in
Fig.~\ref{fig:check_mo_jing_borner}.  For an imposed $T-\rho$ relation
with $(T_0,\gamma)=(25{\rm K},1)$, the CDM FPS does show a cut-off at
small scales, but the value of $k_{\rm max}$ increases with increasing
particle count, $N$. The value of $k_{\rm max}$ for $N=896^3$ and
$N=1024^3$ is nearly identical (see
Fig.~\ref{fig:check_mo_jing_borner}).  We run our main analysis with
the box size $L=20\,{\rm Mpc/}h$ and $N=1024^3$ of both DM and gas
particles, the corresponding scale $k_{\rm max, sim}$ is therefore
much larger than $k_s$.

Our resolution is higher than used previously \citep{Viel13} as the
latter work was interested in hotter thermal histories -- IGM with the
temperature $T_0\sim 10000 - 20000\, {\rm K}$ with a non-negligible
thermal smoothing. Note that \citet{Viel13} also recongnized that
$N=512^3$ with $L=20$~Mpc/$h$ resolution is insufficient, but they
applied a correcting factor to all power spectra. This factor was
calibrated with a single simulation with $N=896^3, L=20~{\rm
  Mpc/}h$. We instead rely on the intrinsic convergence of our
simulations in the range of available data.

\section{The flux power spectrum in CDM and WDM}
\label{sec:thermal_history}

\subsection{Varying the cut-off in the FPS}
\begin{figure*}
\centering	
\includegraphics[width=\textwidth]{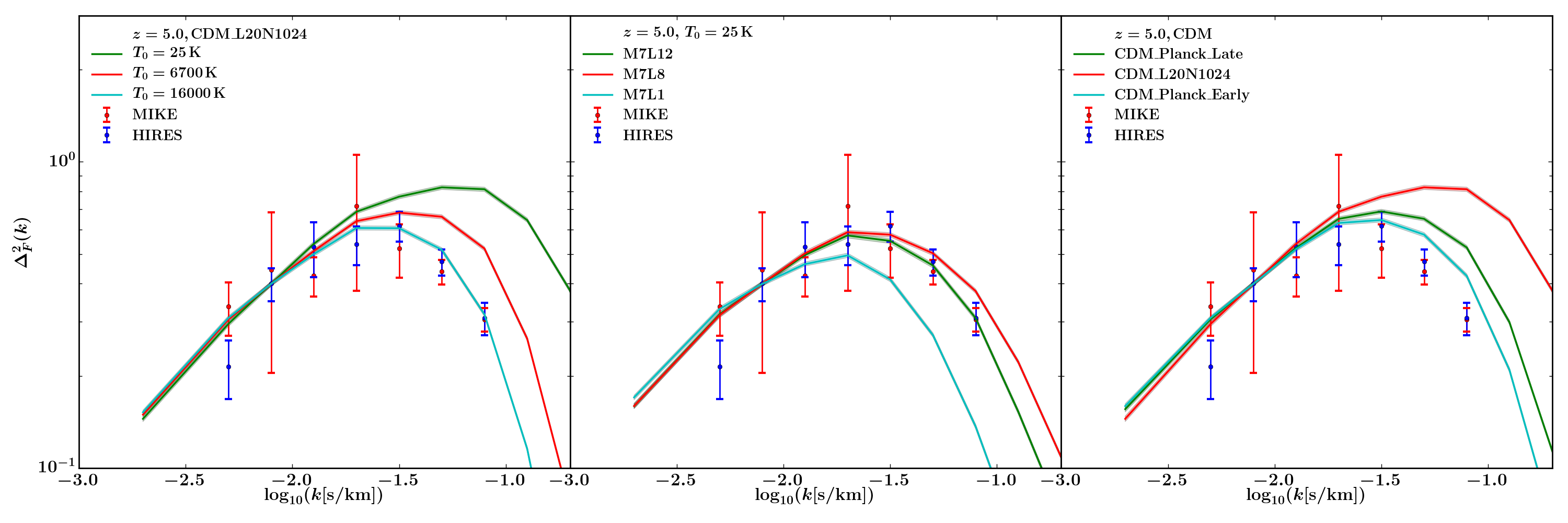}
\includegraphics[width=\textwidth]{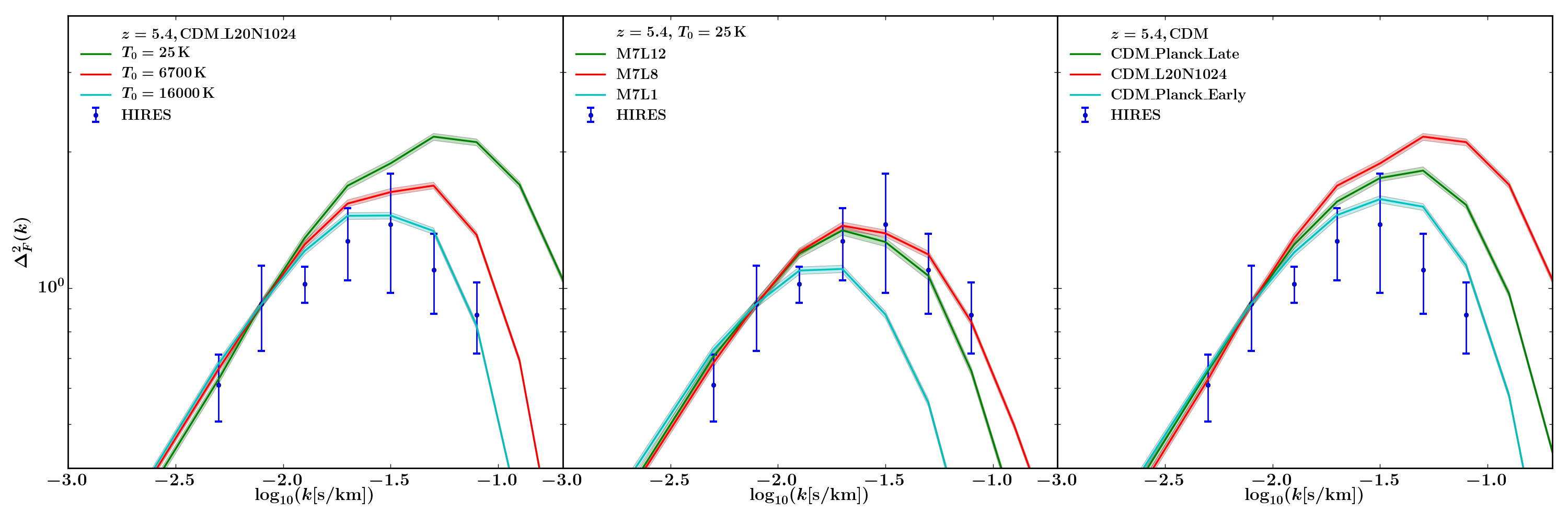}
\caption{ The cut-off in the mock flux power spectrum for various
  models, compared to the {\sc hires} ({\em blue dots with error
    bars}) and {\sc mike} data ({\em red dots with error bars}) at
  redshifts $z = 5.0$ (upper panels) and $z=5.4$ (lower panels).  For
  illustration purposes, we have scaled the amplitudes of the mock FPS
  in all cases such that it agrees with the {\sc hires} value for the
  second point from the left, as a result different FPS in the same
  panel have different $\tau_{\rm eff}$.  {\em Left panels}: model
  {\tt CDM\_L20N1024} with three imposed temperature-density relations
  for $T_0=16000$~K, 6700~K, and 25~K ({\em cyan, red and green
    curves, respectively}).  Doppler broadening introduces a cut-off
  in the FPS that resembles the observed cut-off, for temperatures
  $\sim 2\times 10^4$~K.  {\em Middle panels:} WDM simulations {\tt
    WDM\_L1}, {\tt WDM\_L8} and {\tt WDM\_L12} ({\em cyan, red and
    green curves, respectively}), with negligible Doppler broadening,
  $T_0=25$~K. DM free-streaming alone produces a cut-off in the FPS
  that resembles the observed cut-off for $L_6 =8$ and $12$.  {\em
    Right panels:} CDM simulations {\tt CDM\_L20N1024} without
  pressure effects ({\em red}) compared to the simulations where the
  pressure effects are modeled using the reionisation model of
  \protect\citet{onorbe2016} : late reionisation model in {\tt
    CDM\_Planck\_Late} (green curve) and early reionisation model in
                {\tt CDM\_Planck\_Early} (cyan curve).  To illustrate
                the effects of pressure history alone, the Doppler
                broadening of the lines is reduced by assigning the
                uniform temperature of $T_0=25$~K in post-processing.}
\label{fig:cutoff}
\end{figure*}
We begin this section with illustrating how Doppler broadening, WDM
free-streaming, and pressure smoothing, as quantified by $\lambda_b$,
$\lambda_{\rm DM}$ and $\lambda_p$, respectively, all lead to cut-off
in mock FPS.  Our results are summarized in Figure~\ref{fig:cutoff}.

Doppler broadening introduces a cut-off in the FPS, which in the case
of CDM, resembles the observed cut-off for an imposed power-law
temperature-density relation~\eqref{eq:TDR}, with $T_0\sim 2\times
10^4$~K and $\gamma=1$, as shown in the left panels of
Fig.~\ref{fig:cutoff}, see also \cite{onorbe2017}.
	
Even in the absence of Doppler broadening, WDM free-streaming
introduces a cut-off in the FPS which resembles the observed cut-off
for sufficiently \lq cold\rq\ WDM models. Those with Lepton asymmetry
parameter $L_6=8$ or 12, middle panel of Fig.~\ref{fig:cutoff}, appear
consistent with the {\sc hires} data. (We will perform a more detailed
statistical comparison below.)

Finally the right panel in Fig.~\ref{fig:cutoff} shows the effects of
pressure smoothing on the cut-off in the CDM case.
	
\subsection{Comparison between mock and observer FPS cut-off}\label{sec:chi2}
We have varied the parameters of our models to obtain the best fit to
the cut-off in the FPS by performing a $\chi^2$ analysis.  To this end
we use the evolution of the photo-ionisation and photo-heating rate of
the \emph{LateR} reionization model of \citet{onorbe2016}, impose the
temperature-density relation with $\gamma=1$ in post-processing, and
scale the simulated mean transmission to a range of values
characterised by $\tau_{\rm eff}\equiv -\log\langle F\rangle$.  As
described in Section~\ref{sec:mock}, we convolve the mock spectra with
a Gaussian to mimic instrumental broadening, rebin to the pixel size
of the spectrograph, and add Gaussian noise with standard deviation
independent of wavelength and flux, corresponding to a signal to noise
of 15 at the continuum level.  We compute a grid of mock FPS, varying
$T_0$ and $\tau_{\rm eff}$ for CDM and WDM models.  We compare the
mock FPS to the observed FPS at redshifts $z=5$ and $z=5.4$.  When
doing the comparison we take into account that the scattering between
different realisations is large due to the small size of QSO samples
(see Section~\ref{sec:data} for details). We take into account the
sample variance by computing the $\chi^2$ of a model using the
covariance matrix computed from {\tt EAGLE\_REF} (as the boxsize of
our reference simulation is not large enough to compute the covariance
matrix).  The rationale behind choosing {\tt EAGLE\_REF} was its large
boxsize.  the total length of the lines-of-sight in simulation was
chosen equal to the total length of the observed QSO sample for each
redshift range. Although EAGLE simulations does not have sufficient
resolution at the smallest scales, we expect that the covariance is
reproduced correctly.

\begin{figure*}
  \includegraphics[width=\columnwidth]{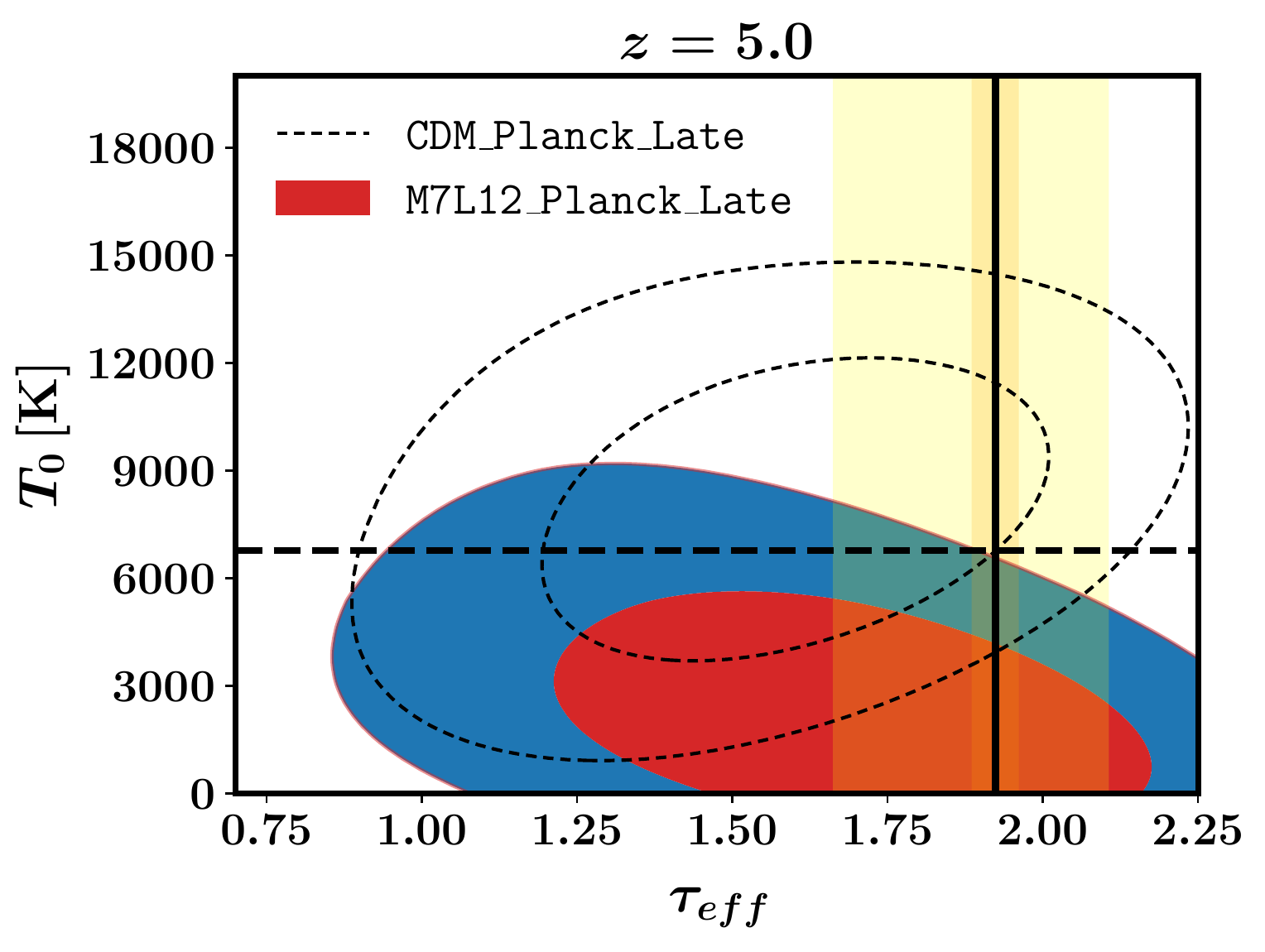}
  \includegraphics[width=0.98\columnwidth]{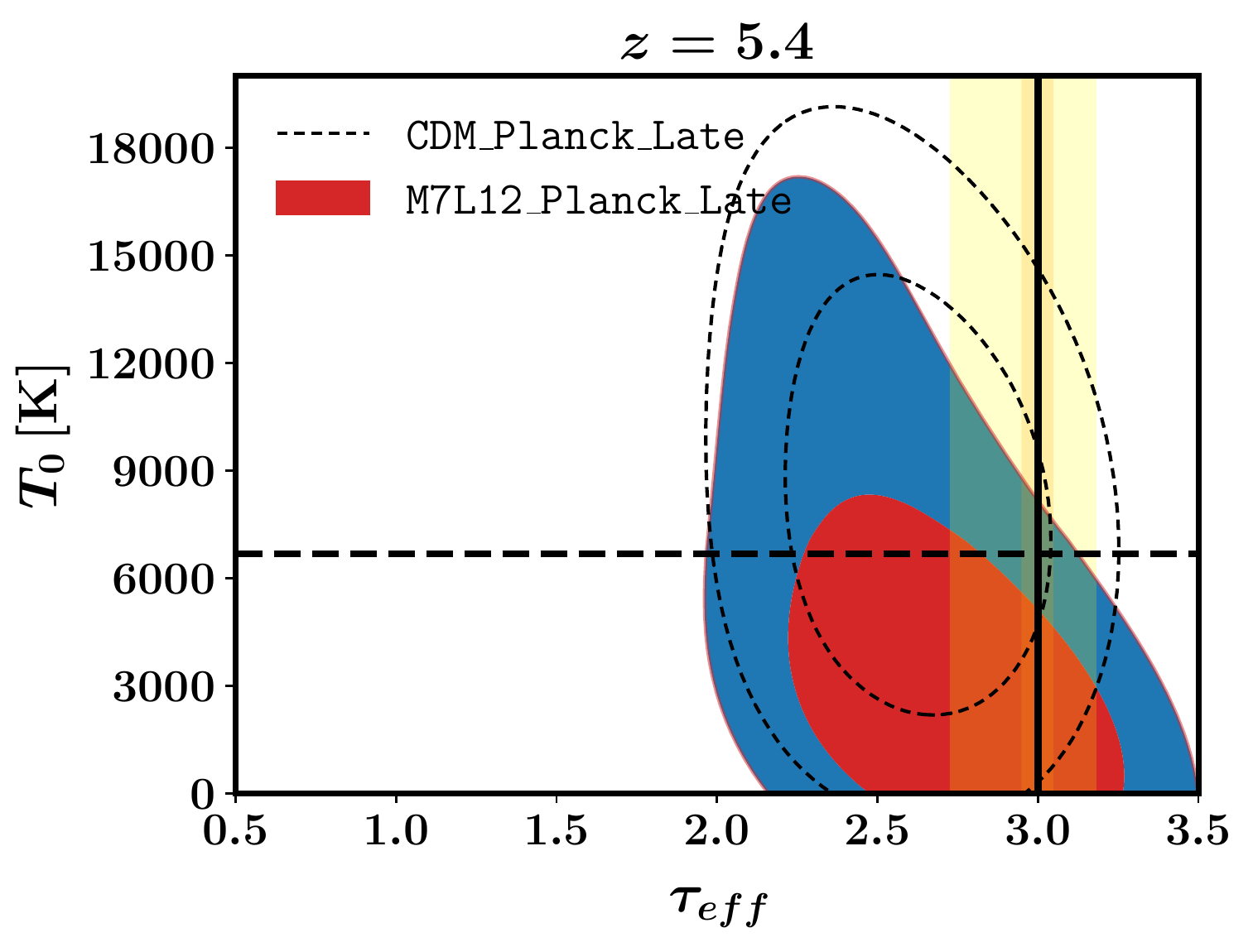}
  \caption{Confidence levels of mock FPS compared to the observed FPS
    of {\sc hires} for redshifts $z=5$ (left) and $z=5.4$ (right).  We
    vary the temperature at the mean density, $T_0$, keeping
    $\gamma=1$, and the value of the effective optical depth
    $\tau_{\rm eff}$.  Solid lines and colour shaded areas correspond
    to $68\%$ and $95\%$ uncertainty intervals for the $m_{\rm
      SN}=7\,{\rm keV}$ and $L_6=12$ WDM model, dashed lines are the
    same for the CDM model.  Both models used the late reionisation
    model {\it LateR} from \protect\cite{onorbe2016}.  The contours
    take into account both {\sc hires} error bars as reported by
    \citet{Viel13} and additional errors due to finite number of
    quasars in the dataset.  The black solid vertical line is the
    directly estimated $\tau_{\rm eff}$ as reported in
    \protect\citet{Viel13}.  The horizontal line shows the value of
    $T_0$ as obtained in simulations with {\it LateR} UVB and without
    post-processing.  It is in full agreement with the results of
    \citet{onorbe2016}.  The systematic uncertainty on $\tau_{\rm
      eff}$ coming from the sample variance is estimated to be $\sim
    10\%$, and we have indicated the resulting uncertainty on
    $\tau_{\rm eff}$ with the orange shade.  The uncertainty on
    $\langle F \rangle$ due to continuum fitting is reported to be at
    the level $\sim 20\%$, and we have indicated the resulting
    uncertainty on $\tau_{\rm eff}$ with the yellow shade.
    \label{fig:bichi2}}
\end{figure*}

The resulting contours for $68\%$ and $95\%$ confidence levels for
{\sc hires} data are shown in Fig.~\ref{fig:bichi2}. In
Table~\ref{tab:bestfit} we have compiled the values of  the $\chi^2$
for the best-fitting models.
\begin{table}
\centering
\caption{Values of $\chi^2$ for the  best-fitting models shown in
  Figure~\ref{fig:bichi2}. The number of dof is 5.\label{tab:bestfit}}
\begin{tabular}{l|c|c}
  \hline
  model &$z$ & $\chi^2$ \\
  \hline
  CDM\_Planck\_Late & 5.0 & 2.20 \\
      & 5.4 & 3.25 \\
  \hline
  M7L12\_Planck\_Late  & 5.0 & 3.44 \\
      & 5.4 & 2.85 
\end{tabular}
\end{table}

 For completeness, in Appendix~\ref{app:allz} we have shown the same
 analysis for the HIRES data-sets at the redshift intervals centered
 on $z=4.2$ and $z=4.6$, that have already been discussed in
 \citep{Viel13}.

\begin{figure*}
\centering	
\includegraphics[width=\columnwidth]{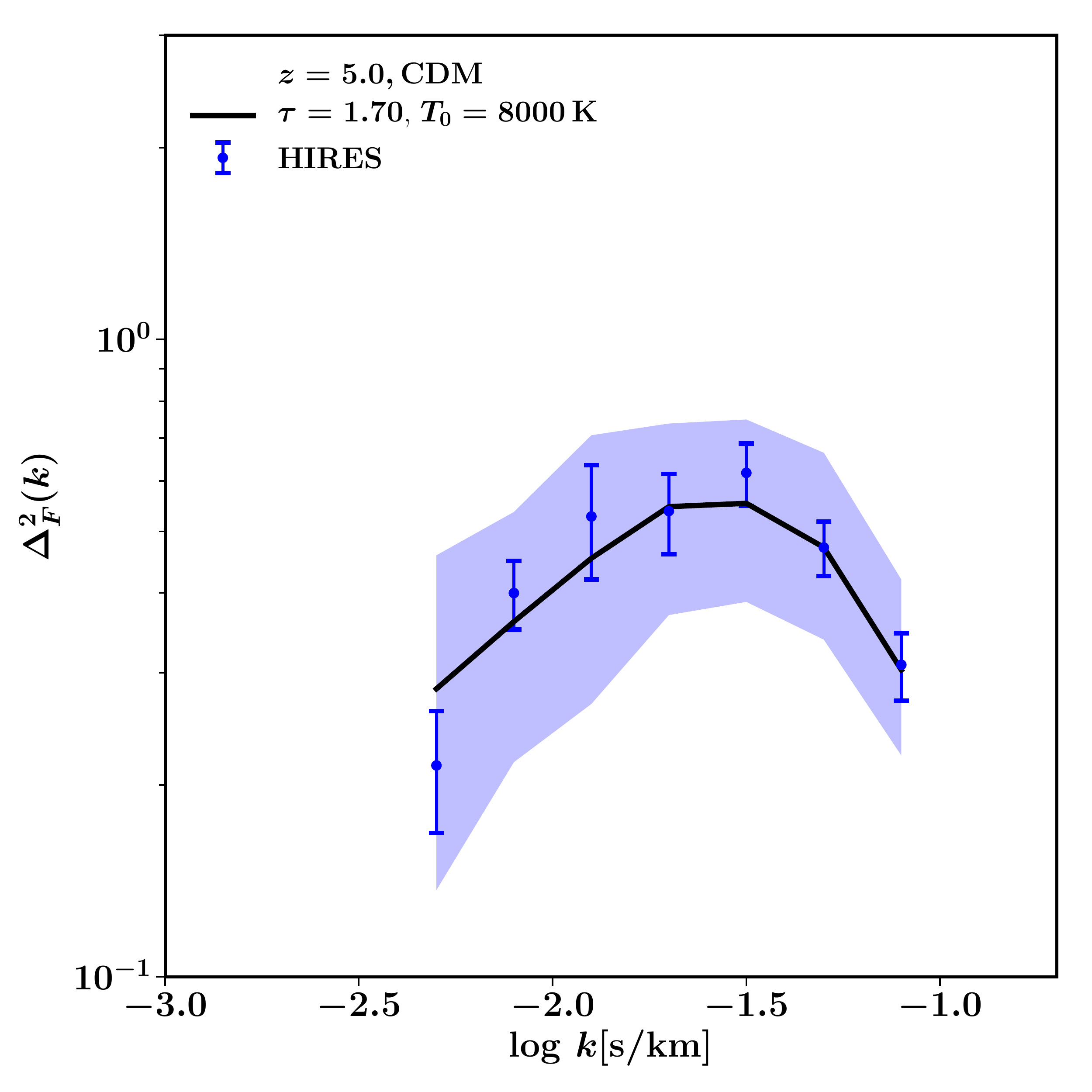}~
\includegraphics[width=\columnwidth]{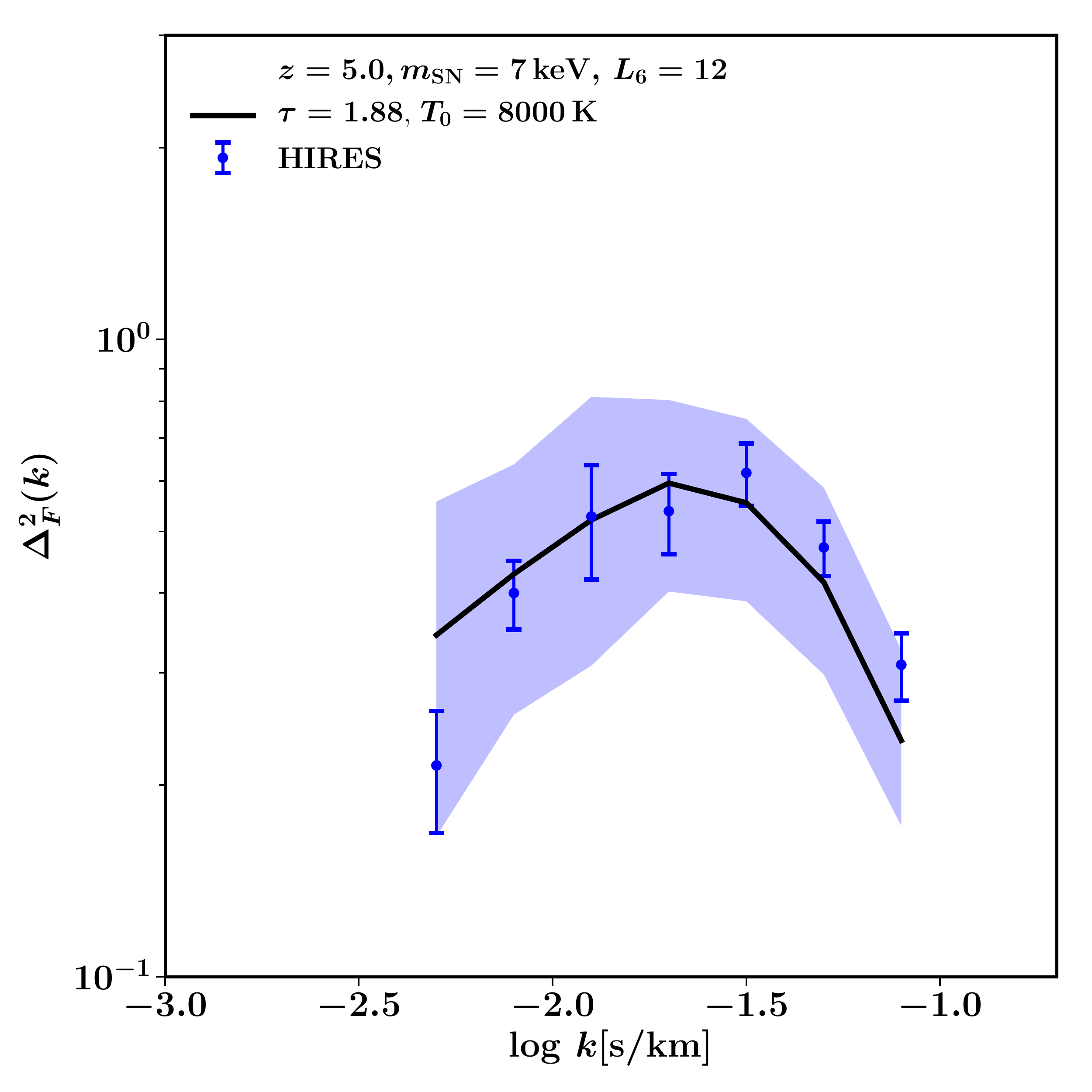}
\\\includegraphics[width=\columnwidth]{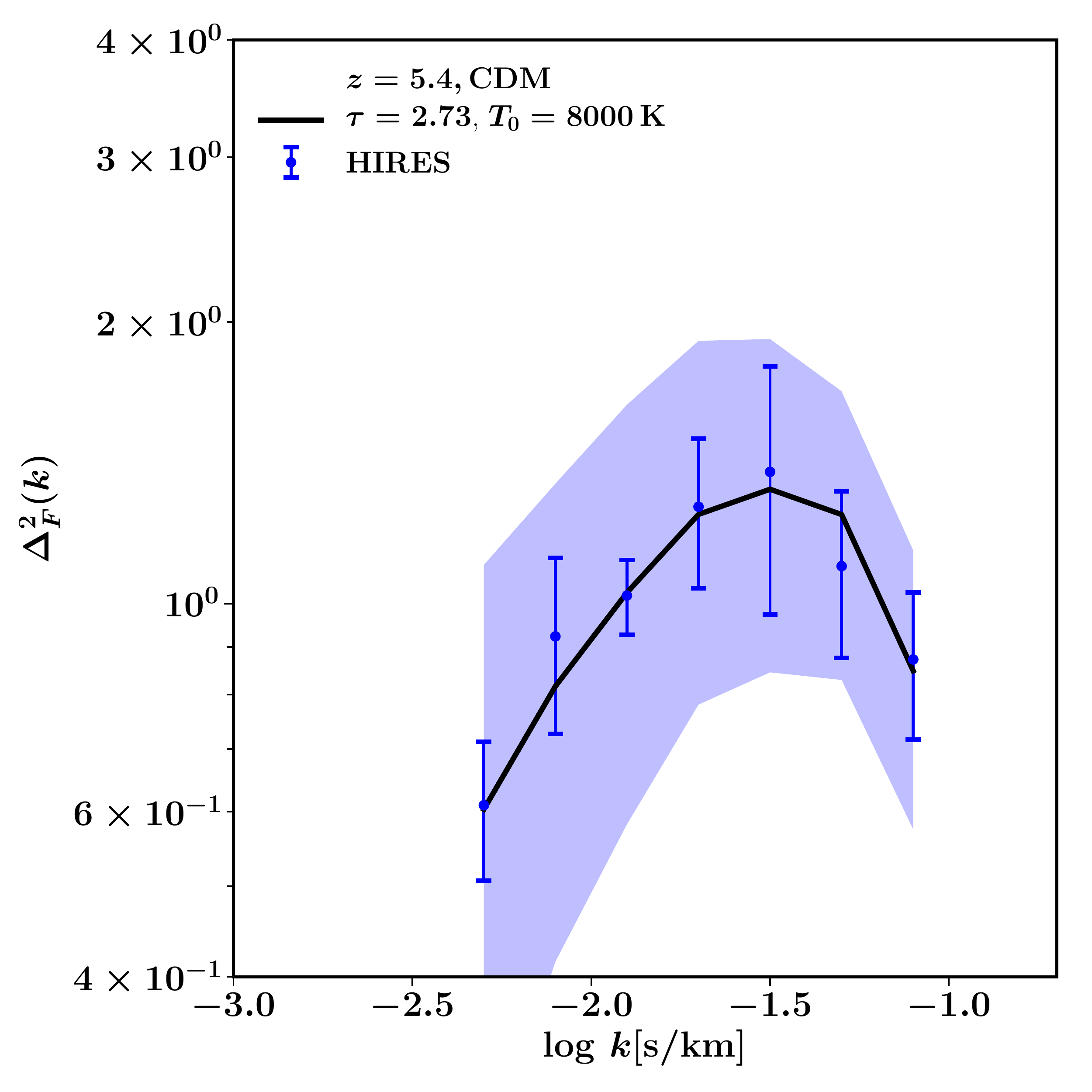}~
\includegraphics[width=\columnwidth]{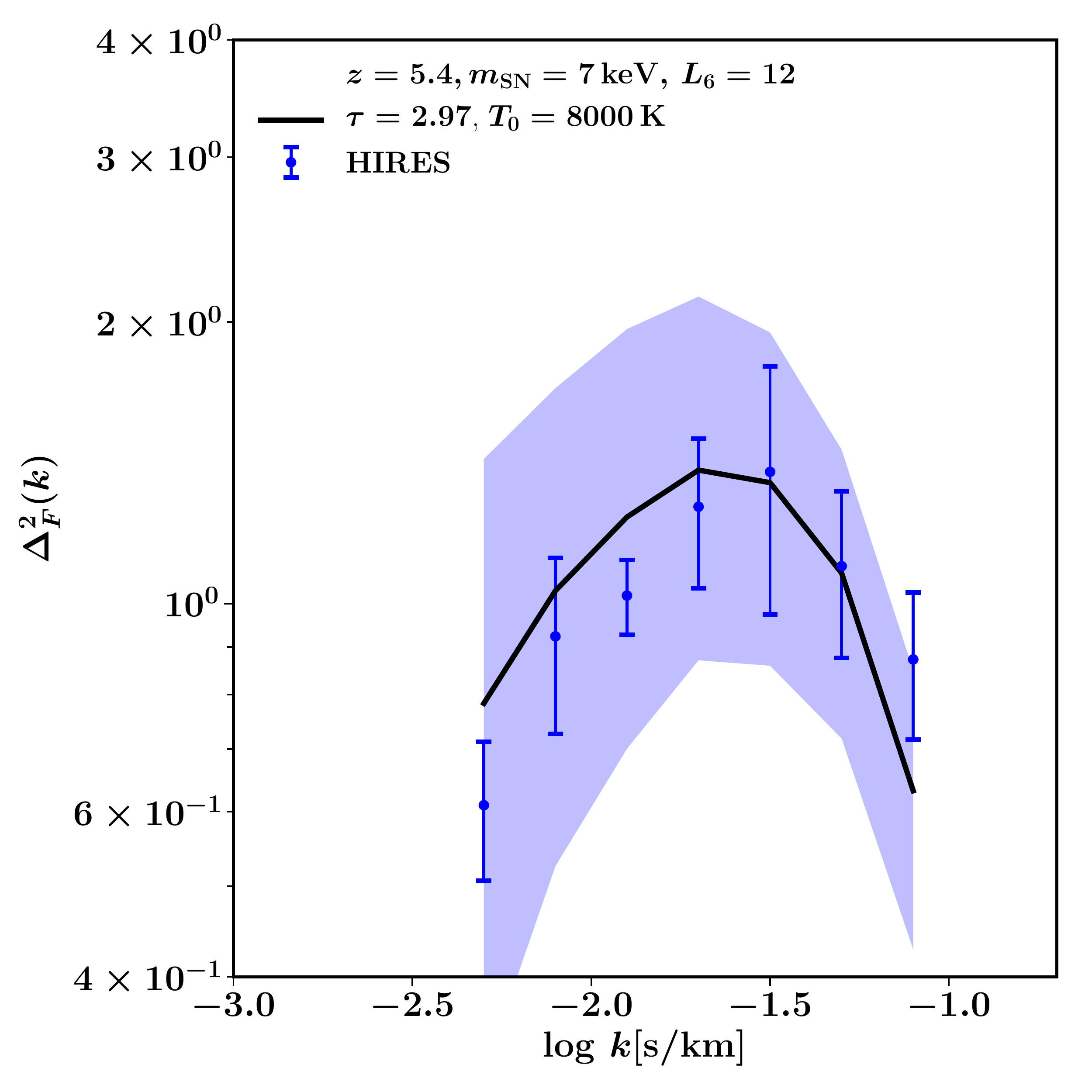}
\caption{ Examples of CDM and WDM models with realistic thermal
  histories, consistent with the high-resolution \lya\ data.  For both
  models we choose $T_0 = 8000$~K as predicted by our simulations with
  \textit{LateR} UVB from \citet{onorbe2016}.  The observed FPS
  inferred from {\sc hires} is plotted as blue symbols with error bars
  as reported by \protect\citet{Viel13}.  One should keep in mind that
  the data points are correlated and therefore do not fluctuate
  independently.  Shaded regions around the model show the variance
  due to different realisations of mock FPS (with the total length of
  the lines-of-sight in simulations equal to the length of observed
  spectra in the dataset for each redshift interval).  The mock
  spectra have best-fit effective optical depth $\tau_{\rm eff} \equiv
  -\log\left<F\right>$ for the fixed uniform temperature $T$ imposed
  in post-processing.  Top panels are for redshift $z=5$ and bottom
  panels -- for redshift $z=5.4$ for CDM ({\em left panels}) and M7L12
  SN model ({\em right panels}).  The simulations are {\tt
    CDM\_Planck\_Late} (left panels) and {\tt M7L12\_Planck\_Late}
  (right panels).
\label{fig:best_fit}}
\end{figure*}

{ As can be seen already from Fig.~\ref{fig:cutoff} (central panel),
  the WDM model M7L12 has the FPS suppression due to the
  free-streaming that is consistent with the data.  Therefore when
  varying $T_0$ in post-processing, WDM prefers temperatures with the
  scale $\lambda_b \ll \lambda_{\rm DM}$, see Fig.~\ref{fig:bichi2}.
  At the same time, our simulation \texttt{M7L12\_Planck\_Late}
  predicts a temperature $T_0^{\rm sim} \simeq 7700$~K at both
  redshifts $5.0$ and $5.4$ (also in agreement with findings of
  \citet{onorbe2016}) .  From Fig.~\ref{fig:bichi2} we see that the
        {\sc hires} data is consistent with $T_0^{\rm sim}$ within its
        $95\%$ confidence interval.  Thus our procedure of
        post-processing is self-consistent -- the temperature
        predicted by the simulations is consistent with the data.  We
        show in Fig.~\ref{fig:best_fit} WDM model with this $T_0$~K as
        an example of a model with realistic thermal history,
        compatible with the data A proper analysis, that varies all
        three scales: $\lambda_p$, $\lambda_b$ and $\lambda_{\rm DM}$
        will be done elsewhere.  }
\section{Discussion}
\label{sec:discussion}
\begin{figure}
  \centering 
  \includegraphics[width=\columnwidth]{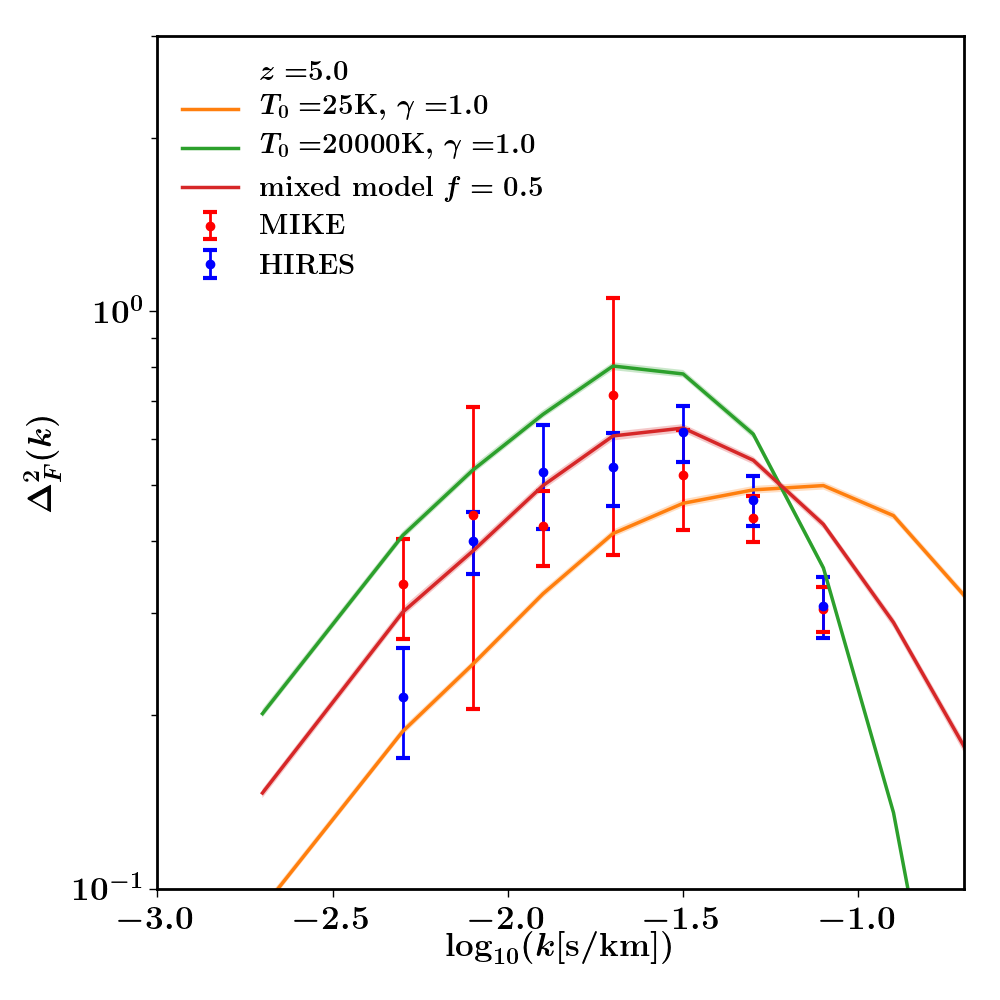}
  \caption{The effect of temperature fluctuations on FPS, for the case that
    bubbles are much larger than our simulated volume at $z=5.0$. This case
    corresponds to a mixing fraction $f=0.5$. For reference, we have drawn the
    data points of the MIKE and HIRES samples.}
  \label{fig:fluct_temp}
\end{figure}
In the previous section we demonstrated that a WDM model with $m_{\rm
  SN}c^2=7$~keV and Lepton asymmetry parameter $L_6\sim 12$ fits the
\lya\ flux power spectrum at redshifts $z=5$ and 5.4 as well as a CDM
model, provided that the Doppler broadening $\lambda_b$ and the
pressure broadening $\lambda_p$ are both sufficiently small. What is
currently {\em known} about these $\lambda$'s?

Since $\lambda_b$ is set by $T_0$, we start by examining limits on the
IGM temperature. When neutral gas is overrun with an ionisation front
during reionisation, the difference between the energy of the ionising
photon and the binding energy of \ion{H}{I}, $\Delta E=h\nu-13.6~{\rm
  eV}$, heats the gas. In the case of \ion{H}{II} regions, gas will
also cool through line excitation and collisional cooling, resulting
in a temperature immediately following reionisation of $T_{0,{\rm
    reion}}\leqslant 1.5\times 10^4$~K \citep{Miralda00,Miralda94}. In
the case of reionisation, the low density of the IGM suppresses such
in-front cooling, and the numerical calculations of \cite{McQuinn12}
suggest $T_{0,{\rm reion}}=1-4\times 10^4$~K, depending on the
spectral slope of the ionising radiation. Following reionisation, the
IGM cools adiabatically while being photoheated, preserving some
memory of its reionisation history \citep{Theuns02,Hui03}. Therefore
the value of $T_0$ at $z=5.4$ is set by $T_{0,{\rm reion}}$, the
redshift $z_{\rm reion}$ when reionisation happened, and the shape of
the ionising radiation that photoheats the gas subsequently. For $T_0$
to be sufficiently low then requires that $T_{0,{\rm reion}}$ is low,
that $z_{\rm reion}\gg 5.4$, and that the ionising radiation is
sufficiently soft.

Taking $z_{\rm reion}=7.82$ from \cite{Planck18} and $T_{0,{\rm
    reion}}=1.5\times 10^4$~K yields a guesstimate for the lower limit
of $T_0\sim 0.8\times 10^4$~K at $z=5.4$, consistent with the value of
$T_0\sim 10^4~$K suggested by \cite{onorbe2017} that we used in the
previous section. There is now good evidence that \ion{He}{II}
reionised at $z~\sim 3.5$, much later than \ion{H}{I} and \ion{He}{I}
\citep{Jakobsen94,Schaye00,LaPlante17,Syphers14}, as the ionising
background hardens due the increased contribution from quasars.  This
suggests that the ionising background during reionisation was unable
to ionise \ion{He}{II} significantly and hence was relatively soft. So
conditions for low $T_0$ seem mostly satisfied.

However the FPS also depends on the slope $\gamma$ of the
temperature-density relation, not just $T_0$. As gas is impulsively
heated during reionisation, the heat input per hydrogen atom is mostly
independent of density, driving $\gamma\rightarrow 1$. The heating
rate then drops as the gas becomes ionised, but more so at low density
than at high density. This steepens the TDR asymptotically to
$\gamma-1=1/(1+0.7)\sim 0.6$, with the factor 0.7 resulting from the
temperature dependence of the {\sc case-A} \ion{H}{II} recombination
coefficient \citep{Theuns98,Upton16}. The characteristic time-scale
for approaching the asymptotic value is of the order of the Hubble
time. If reionisation indeed happens late, $z\sim 7.5$, then we would
expect $1<\gamma< 1.6$.

Observationally, the IGM temperature is constrained to be at the level
$T_0 \gtrsim 8000$~K at $z \lesssim
4.6$~\citep{Schaye00,McDonald:2000nn,Lidz:2009ca,becker2011} (see
e.g. \cite{Upton16} for a recent discussion). At $z\approx 6.0$ there
is a single measurement in the near zone of a quasar that yields $5000
< T_0 <10000$~K (68\% CL, \citet{bolton2012}). Fundamentally, all of
the techniques used to infer $T_0$ observationally are based on
identifying and computing the statistics of sharp features in
\lya\ forest spectra, and comparing these to simulated spectra. This
implies that the $T_0$ inferred implicitly depends on $\lambda_{\rm
  DM}$.

Combining the theoretical prejudice and the measurements, we conclude
that a value of $T_0\sim 8000$~K or even colder at redshifts around 5
is not unreasonable and definitely not ruled out. Using
Eq.~(\ref{eq:kmaxb}), such a value of $T_0$ yields $k_{{\rm
    max},b}=0.12$~(km~s$^{-1}$)$^{-1}$.

What do we know about $\lambda_p$? From a theoretical perspective,
this \lq Jeans\rq\ or \lq pressure broadening\rq\ results from Hubble
expansion over the finite extent of the absorbing filament
\citep{Garzilli15}. In the linear approximation, this results in a
value of $\lambda_p$ that is in general smaller than the Jeans length
$\lambda_J$ because gas needs to physically expand away from the much
thinner dark matter filaments before it reaches the final filament
width \cite[]{Gnedin98}. In the special case but not unrealistic case
where $T_0\approx 0$ before reionisation and a constant after
reionisation, \cite{Gnedin98} find
\begin{equation}
\left({\lambda_p\over\lambda_J}\right)^2={3\over 10}\left[1+4\left({1+z\over 1+z_{\rm reion}}\right)^{5/2}-5\left({1+z\over 1+z_{\rm reion}}\right)^2\right]\,.
\end{equation}
Taking again $z_{\rm reion}=7.82$ yields $\lambda_p=(0.3-0.2)\lambda_J$ at redshift $z=5$ and 5.4, respectively, or in terms of the cut-off in the FPS using Eq.~(\ref{eq:kmaxp}), $k_{{\rm max},p}=0.3-0.4$~(km~s$^{-1}$)$^{-1}$.

Comparing these estimates of $\lambda_b=1/k_{{\rm max},b}\sim 8~{\rm
  km~s}^{-1}$ and $\lambda_p=1/k_{{\rm max},p}\sim 3~{\rm km~s}^{-1}$,
it is not surprising that WDM free-streaming with $\lambda_{\rm
  DM}\sim 10~{\rm km~s}^{-1}$ dominates the cut-off in the FPS in the
WDM case. Since this is also close to the observed cut-off scale show
why such a WDM model is consistent with the data. We note in passing
that a small value of $\lambda_b$ favours reionisation to be early,
whereas a small value of $\lambda_p/\lambda_J$ favours reionisation to
be late. The current value of $z_{\rm reion}\approx 7.8$ happens to be
a good compromise between the two.

The plausible patchiness of reionisation introduces complications. For
example the large-scale amplitude of the FPS may be more a measure of
the scale and amplitude of temperature fluctuations or of fluctuations
in the mean neutral fraction, rather than being solely due to density
fluctuations that we simulate. If that were the case, then our
simulations should not match the measured FPS on large-scales, since
we have not included these effects (see
e.g. \citet{Becker:2014oga}). Furthermore, what is the meaning of
$\lambda_b$ or $\lambda_p$ in such a scenario, where these quantities
are likely to vary spatially? Matching the cut-off in the FPS might
pick-out in particular those regions where both $\lambda_b$ and
$\lambda_p$ are unusually small.

To illustrate the effect of fluctuations on the FPS, we contrast the
FPS of two sets of mock spectra with an imposed temperature-density
relation with different values of $T_0$: 25~K ({\em i.e.} negligible
Doppler broadening and $T_0=2\times 10^4$~K in
Fig.~\ref{fig:fluct_temp}, as well as a mock sample that uses half of
the spectra from each of the two models. The FPS for the
single-temperature models are normalized to have the same mean
effective optical depth, $\tau_{\rm eff}=2.0$, the mixed-temperature
model is computed from the two normalized single-temperature models,
and it is not normalized further.  We find that in the mixed model the
FPS is intermediate between the FPS of the hot and cold models. Hence,
if the hot model represents the recently reionized regions in the IGM
and the cold model the patches that were reionized previously and then
cooled down, the mixed model looks like a model that is colder than
the regions in the IGM that were reionized more recently.

Fig.~\ref{fig:fluct_temp} illustrates that fluctuations essentially
decouple the behaviour of the FPS at large and small scales. If this
is the case of the real IGM, then what we determine to be $T_0$ from
fitting the cut-off does not correspond to either the hot or the cold
temperature.  We leave a more detailed investigation of patchiness on
the FPS and how that impacts on constraints on $\lambda_{\rm DM}$ to
future work.

\section{Conclusions}
\label{sec:conclusions}

The power spectrum of the transmission in the \lya\ forest (the flux
power spectrum, FPS), exhibits a suppression of power on scales
smaller than $\lambda_{\rm min}=1/k_{\rm max}\sim 30~{\rm
  km~s}^{-1}$. Several physical effects may contribute to this
observed cut-off: ({\em i}) Doppler broadening resulting from the
finite temperature $T_0$ of the intergalactic medium (IGM), ({\em ii})
Jeans smoothing due to the finite pressure of the gas, and ({\em iii})
dark matter free streaming; these suppress power below scales
$\lambda_b$, $\lambda_p$ and $\lambda_{\rm DM}$, respectively. We have
shown in Section~3 that, when $\lambda$ is expressed in velocity
units, $\lambda_b$ and $\lambda_p$ are independent of redshift $z$ for
a given value of $T_0$, whereas $\lambda_{\rm DM}\propto
(1+z)^{1/2}$. This means that any smoothing of the density field due
to warm dark matter (WDM) free-streaming will be most easily
observable at high-redshift, and the observed FPS may provide
constraints on the nature of the dark matter
\citep{Viel13,irsic2017,Irsic:2017yje,Murgia:2018now}, and possible be
a \lq WDM smoking gun\rq.

{\noindent  In this paper we tried to answer two questions:
\begin{compactitem}[--]
\item Does the observed cut-off in the FPS favour cold or warm dark
  matter, or can both models provide acceptabale fits to the existing
  data?
\item Are the WDM models with large $\lambda_{\rm DM}$ that were previously excluded allowed if one considers a less restrictive thermal history?
\end{compactitem}

To answer these questions we run a set of cosmological hydrodynamical
simulations at very high resolution, varying $\lambda_b$, $\lambda_p$
and $\lambda_{\rm DM}$ independently. We then compute mock spectra
that mimic observational limitations (noise, finite spectral
resolution and finite sample size), and compare the mock FPS to the
observed FPS.

We demonstrate that all three effects ({\em i.e.} Doppler broadening,
Jeans smoothing and DM free-streaming) yield a cut-off in the FPS that
resembles the observed cut-off. Of course in reality all three effects
will contribute at some level. In particular, Doppler broadening and
Jeans smoothing both depend on the temperature $T_0$ of the IGM, and
so always work together.

To answer the two questions posed above, we have tried to fit the
observed FPS at redshifts $z=5$ and 5.4 with ({\em i}) a CDM model
(which has $\lambda_{\rm DM}=0$), varying $T_0$ and the thermal
history, and ({\em ii}) the particular case of a resonantly produced
sterile neutrino WDM model (characterised by the mass of the particle,
$m_{\rm DM}c^2=7$~keV, and the Lepton asymmetry parameter $L_6$,
\citet{Boyarsky:2008mt}), varying $L_6$, $T_0$ and the thermal
history.

In addition to motivations based on particle physics (see {\em e.g.}
\citet{Boyarsky:2018tvu} our particular choice of WDM particle is
motivated by the fact that ({\em i}) its decay may have been observed
as a 3.5~keV X-ray line in galaxies and clusters of galaxies
\citep{Boyarsky:2014jta,Bulbul:2014sua,Boyarsky:2014ska}, ({\em ii})
it produces galactic (sub)structures compatible with
observations~\citep{Lovell16,Lovell17}, and ({\em iii}) it is
apparantly ruled out by the observed FPS ~\citep{baur2017}.

Fig.~\ref{fig:best_fit} shows how the {\sc hires} data is compatible
with CDM and SN cosmologies if we choose relatively late reionisation
model ({\em LateR} of \citet{onorbe2017}) so that $\lambda_p$ is small
and $T_0\approx 7-8000$~K as predicted by this model.  Both the
assumed late reionisation redshift, and the relatively low value of
$T_0$, are reasonable and consistent with expectations and previous
work, as we discuss in detail in the Discussion section. Crucially, a
WDM model with $L_6=12$ and the same late redshift of reionisation
also provides an acceptable fit to the data, provided $T_0\leqslant
7000 $~K. With such a low value of $\lambda_b$ and $\lambda_p$, the
FPS cut-off is mostly due to WDM free streaming.

From this comparison we conclude that the observed suppression in the
FPS can be explained by thermal effects in CDM model but also by the
free-streaming in a WDM model: current data do not strongly favour
either possibility.  We also find a reasonable fit for a WDM model
that was previously ruled out by \citet{Viel13} and
\citep{irsic2017,Irsic:2017yje,Murgia:2018now}.  Our present analysis
differs in a number of ways:
\begin{enumerate}[\bf 1.]
\item We vary the thermal history of the IGM within the allowed observational limits as discussed by~\citet{onorbe2016, onorbe2017}.
The previous works modeled the UVB according to \citet{haardt2001}. The latter scenario is known to reionise the Universe too early with respect to current observations \citep{onorbe2016}, plausibly overestimating $\lambda_p$.

\item We did not use any assumptions about the evolution $T_0(z)$ but inferred ranges of $T_0$ at $z= 5.0$ and $z=5.4$ based on theoretical considerations and limits inferred from the \lya\ data (see also~\citet{Garzilli:2015iwa}). 

\end{enumerate}

We also reconsidered the impact of peculiar velocities (\lq redshift
space distortions\rq), which were claimed to affect the appearance of
a cut-off at the smallest scales \citep{desjacques2004}, but found
these not to be important at the much higher resolution of our
simulations.

We also demonstrated that spatial fluctuations in temperature, which
are expected to be present close to reionisation, may dramatically
affect the FPS. Spatial variations in $T_0$ can dramatically increase
the amplitude of the FPS at the scale of the imposed fluctuations,
effectively decoupling the large-scale and small-scale
FPS. Unfortunately this means that a model without fluctuations in
$T_0$ will yield incorrect constraints on parameters if such
fluctuations {\em are} present in the data. Interestingly, the
nuisance caused by fluctuations in $T_0$ may actually be rather
helpful if the cut-off in the FPS is in fact due to WDM, since in that
case there would be no spatial fluctuations in the location of the
cut-off - and the evolution with redshift of the cut-off would follow
$\lambda_{\rm DM}\propto (1+z)^{1/2}$.

Moving away from \lya\ and studying the small-scale Universe in the
\ion{H}{I} 21-cm line during the \lq Dark Ages\rq\ \citep{Pritchard12}
instead is currently almost science fiction, but ultimately may be the
most convincing way of determining once and for all whether most of
the dark matter in the Universe is warm or cold.

 \section*{acknowledgements}	
This project has received funding from the European Research Council
(ERC) under the European Union's Horizon 2020 research and innovation
programme (ERC Advanced Grant 694896).  TT and CSF CSF were supported
by the Science and Technology Facilities Council (STFC) [grant number
  ST/P000541/1]. CSF acknowledges support from the European Research
Council (ERC) through Advanced Investigator Grant DMIDAS (GA 786910).
This work used the DiRAC Data Centric system at Durham University,
operated by the Institute for Computational Cosmology on behalf of the
STFC DiRAC HPC Facility (www.dirac.ac.uk). This equipment was funded
by BIS National E-infrastructure capital grant ST/K00042X/1, STFC
capital grants ST/H008519/1 and ST/K00087X/1, STFC DiRAC Operations
grant ST/K003267/1 and Durham University. DiRAC is part of the
National E-Infrastructure.

We would like to thank Jose O{\~n}orbe for sharing with us additional
unpublished thermal histories.
\clearpage
\appendix
\section{Convergence of the simulations in
box-size}\label{app:box_convergence}
\begin{figure*}
  \centering 
   \includegraphics[width=\columnwidth]{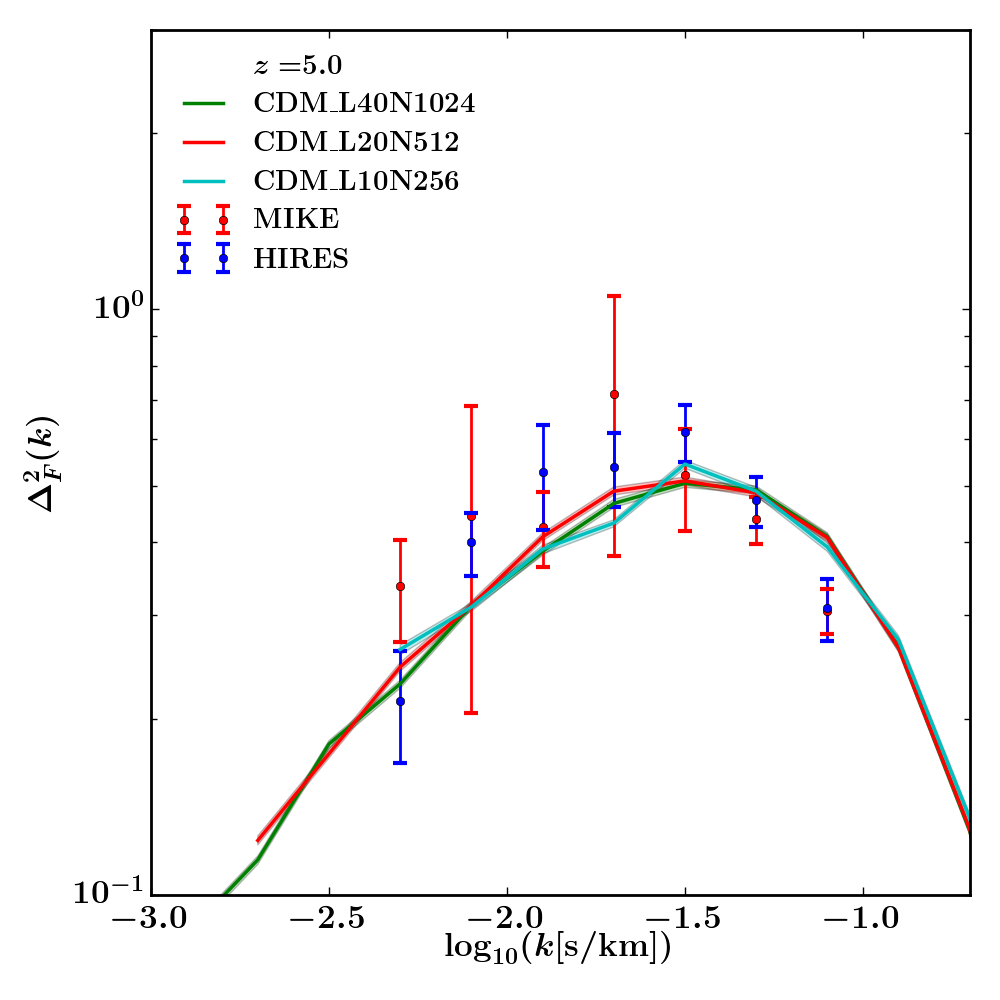} 
   \includegraphics[width=\columnwidth]{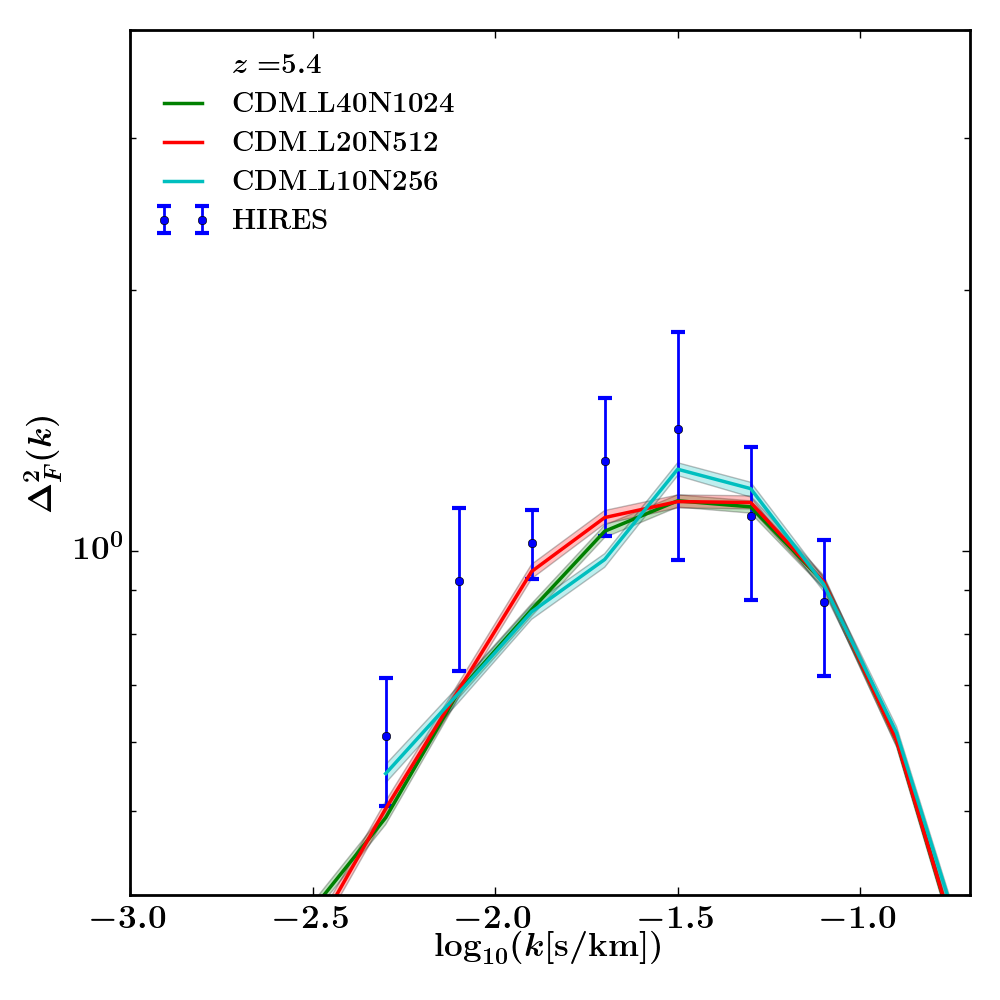} 
  \caption{Study of the box-size needed in the numerical simulations
    to resolve the smallest scales probed by the HIRES and MIKE data
    samples. We show the FPS at $z=5.0$ and $z=5.4$ for three
    simulations without UVB and different box-sizes, yet same
    resolution. We have imposed a uniform temperature $T=25\,{\rm K}$
    in the post-processing of the spectra. We have applied the
    resolution of the HIRES spectrograph to the spectra, but we have
    excluded the effect of noise on the spectra. The FPS are
    normalized to the nominal observed optical depth of the observed
    spectra. The red solid line has a box-size $L=10\,{\rm Mpc}/h$,
    the green solid line $L=20\,{\rm Mpc}/h$, and the orange solid
    line $L=40\,{\rm Mpc/h}$. The FPS for the case of $L=10$,
    $L=20\,{\rm Mpc}/h$, and $L=40\,{\rm Mpc}/h$ agree with each
    other.}
\label{fig:box_convergence}
\end{figure*}

We have investigated the convergence of the FPS in box-size of the
simulation with constant resolution.  In section~\ref{sec:convergence}
we have concluded that we need at least a number of particles
$N=1024^3$ and a boxsize $L=20\,{\rm Mpc}/h$ to resolve the smallest
scales reached by the data. Because we do not have the computing power
to run a simulation with $L=40\,{\rm Mpc}/h$ with this maximal
resolution, we consider three simulations with $L=10,20,40\,{\rm
  Mpc}/h$ and half the resolution. In this limit, we show in
Figure~\ref{fig:box_convergence} that the $L=20\,{\rm Mpc/h}$ is
sufficient to resolve the scales we intend to study.

\section{Effect of peculiar velocities on Flux Power Spectrum}
\label{app:shape}

In CDM cosmologies the real-space MPS, $\Delta^2_{r,3d}(k)$, is a
monotonically increasing function of $k$.  However, in \emph{velocity
  space} over which the FPS observable is built, an additional effect
-- the \emph{redshift space distortions} (RSD) -- affect the shape
\citep{Kaiser:1987qv, Kaiser:1991, McGill:1990, Scoccimarro:2004tg}.
RSD may erase small-scale power in the FPS because peculiar velocities
of baryons are non-zero.

At linear level MPS in velocity space is related to real space by:
\begin{align}
	\Delta^2_{s,3d}(k) =\Delta^2_{r,3d}(k) ( 1+\beta (\vec{k}\cdot\hat{z})^2)^2
\end{align}
where $\hat{z}$ is the direction of observation and constant $\beta$ for linear
scales is given by expression $\delta_{r} = -\beta^{-1}
\vec{\nabla}\cdot\vec{v}$ \citep{Kaiser:1987qv}.

Real-space MPS projected along the line of sight is given by:
\begin{align}
\Delta^2 _ { r,1d } ( q ) &= \frac { q } { 2 \pi } \int d ^{ 2 } k _ { \perp } \frac { \Delta^2 _ { r,3 d } (q ,k _ { \perp } )} { \left( q ^ { 2 } +k _ { \perp } ^ { 2 } \right) ^ { 3 / 2 } }\\
	 &= q \int_q^\infty \frac{dk}{k^2} \Delta^2_{r,3d}(k)
     \label{eq:project3Dto1D}
\end{align}
Clearly, in CDM linear $\Delta^2_{r,1d}(k)$ remains a monotonic
function of $k$.  Non-linear MPS experiences additional growth at
small scales, therefore $\Delta^2_{r,1d}(k)$ does not exhibit a cutoff
also at non-linear level.

Beyond the linear regime it is not possible to compute analytically
the effect of RSD on the MPS. \citeauthor{desjacques2004} have
attempted to address this case, by considering a fitting formula
calibrated to N-body simulations by \citeauthor{mo_jing_borner1997}:
\begin{align}
\Delta^2 _ { s,1d } ( q ) &= q \int _ { q } ^ { \infty } d k \frac { \Delta^2 _ { r,NL } ( k ) } { k ^ { 2 } } \left[ 1 + \beta \left( \frac { q } { k } \right) ^ { 2 } \right] ^ { 2 } D \left[ q \sigma _ { 12 } ( k ) \right]  \\
D \left[ x \right] &= \left[ 1 + \frac { 1 } { 2 } x ^ { 2 } + \eta x ^ { 4 } \right] ^ { - 1 }
\end{align}
where $\sigma_{12}(k)$ is a pairwise velocity dispersion of dark
matter particles, $\Delta^2 _ { s,NL }$ is a nonlinear $3d$ MPS and
$\eta$ is a constant. \citet{desjacques2004} predicted a cut-off on
the scales similar to the cut-off observed in the HIRES and MIKE data.

In order to verify the predictions of \citet{desjacques2004}, we have
performed simulations where thermal effects were switched off,
(Figure~\ref{fig:check_mo_jing_borner}). Obviously, the simulation
results for, \textit{e.g.,}, the IGM temperature are unrealistic in
this case. The purpose of this exercise was to identify the position
of a RSD-induced cut-off, which might have been obscured by thermal
broadening (otherwise it would be have been covered by the cut-off due
to the thermal Doppler effect and cut-off due to the extent of the
structures). We find that the resolution of simulations by
\citeauthor{mo_jing_borner1997} stays significantly below the required
resolution of our convergence analysis: number of particles $N=128^3$
and box-size $L=100\,{\rm Mpc}/h$ \citep{mo_jing_borner1997} against
$N=1024^3,\;L=20\,{\rm Mpc}/h$.  We conclude that the relevant scales
have not been resolved in past simulations.  To support this claim, we
compare the FPS for various resolutions in model cosmologies designed
to remove baryonic effects as much as possible, see
Figure~\ref{fig:check_mo_jing_borner}. Since our high-resolution
simulations exhibit a cut-off at a position $k$'s that is
significantly larger than the reach of the data, we conclude that the
role of RSD in the formation of the cut-off is negligible.

\section{Effect of noise}\label{app:noise}

We investigate the effect of noise on the FPS. In our implementation
of the noise, we have considered a Gaussian noise, with amplitude
independent of flux or wavelength. In a spectrum from a bright quasar,
the S/N is expected to increase with the flux. Because we have
considered a S/N that is constant with flux and matches the S/N
measured at the continuum level, we are likely underestimating the
effect of noise in our analysis.

In some of the previous works on FPS in the Lyman-$\alpha$ forest, in
particular \citep{Viel13,irsic2017}, the effect of noise on the flux
PS is encoded with the application of a correction, that only depends
on the chosen S/N and the redshift (in particular see Figure~16 in
\citet{Viel13}), and not on other parameters of the IGM, such as the
$\tau_{\rm eff}$. We have investigated whether the effect of noise is
independent of the level of ionization of the IGM. In
Figure~\ref{fig:noise} we show explicitly that the ratio between the
FPS computed in the cases with and without noise depends on the value
of the $\tau_{\rm eff}$, and the difference becomes larger on the
smallest scales.  This example was computed for a CDM simulation
without, and with a uniform temperature $T=2\times10^4\,{\rm K}$
imposed in post-processing. The effect of noise on the FPS is
presumably being affected also by the temperature of examined
spectra. Hence, we have resorted to including the effect of noise in
our analysis by applying the noise to the spectra and then computing
the resulting FPS.

\begin{figure*}
  \centering 
   \includegraphics[width=\columnwidth]{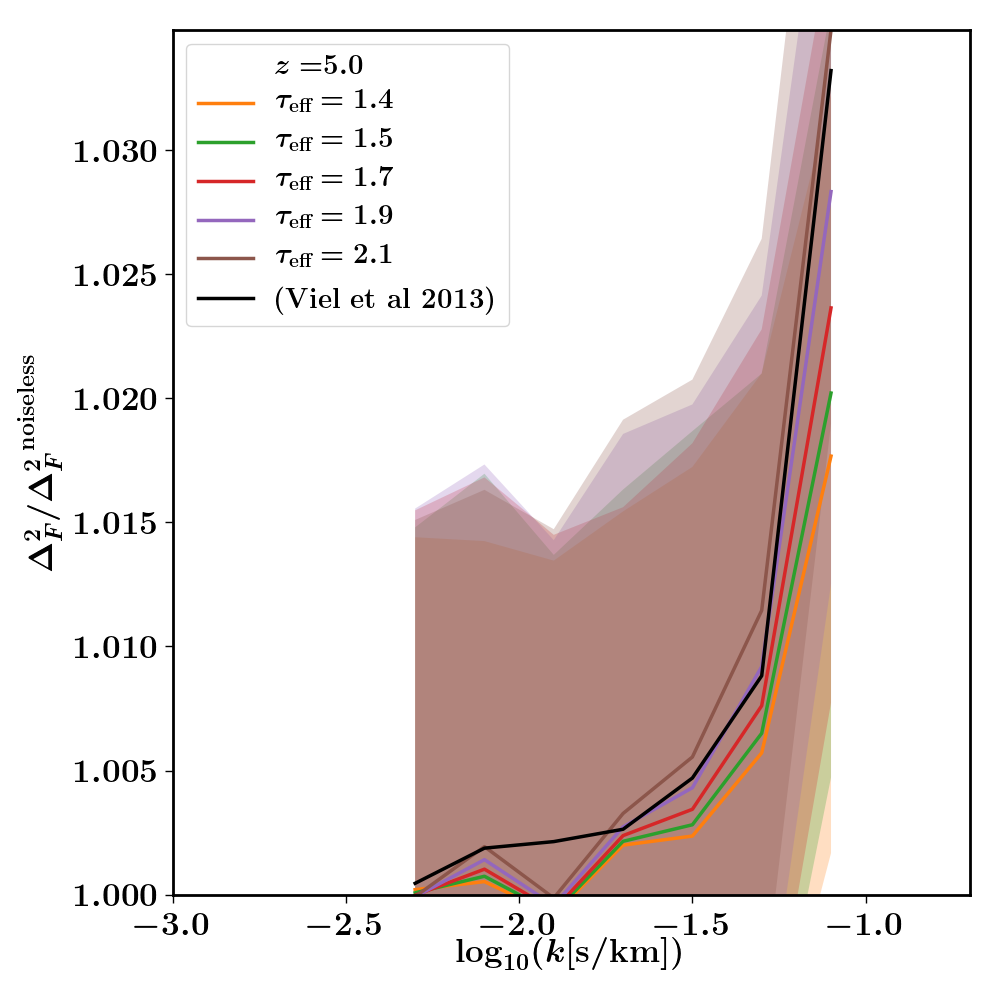} 
   \includegraphics[width=\columnwidth]{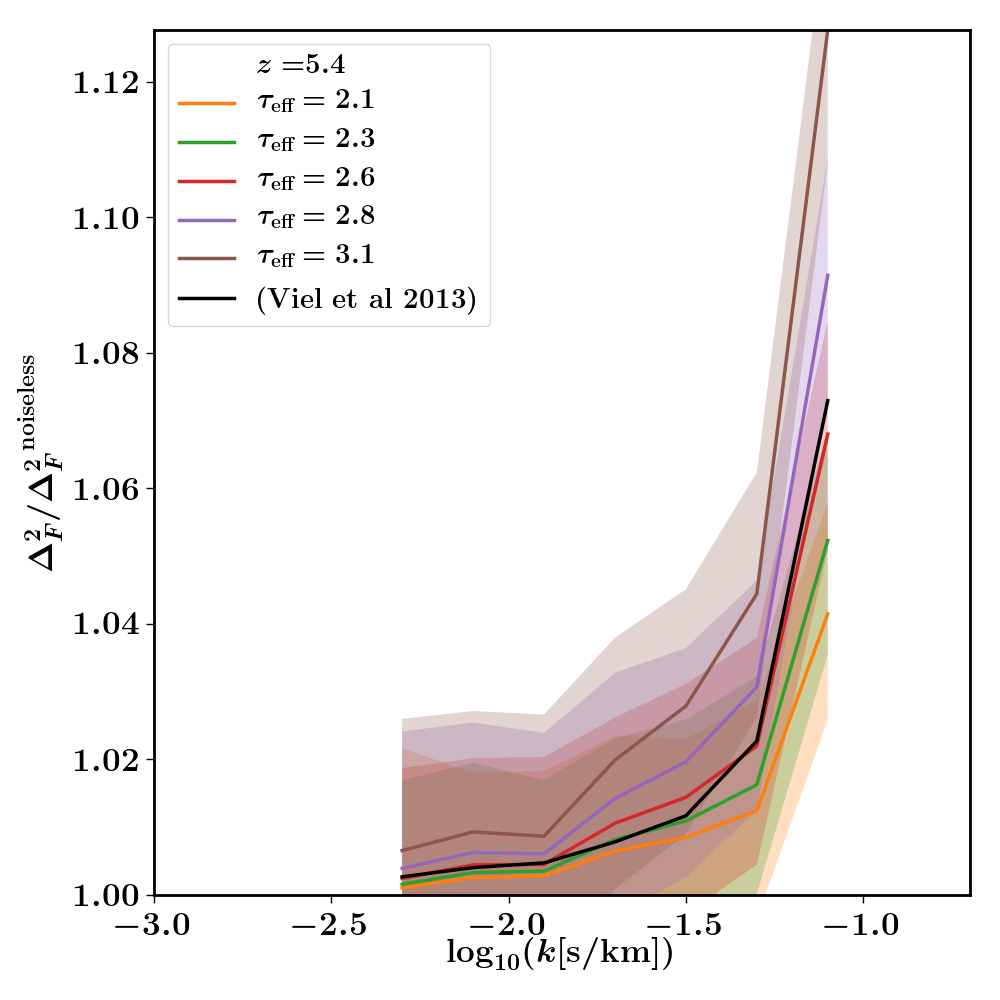} 
  \caption{The effect of noise on FPS and its dependence from
    $\tau_{\rm eff}$. We show the ratio between the FPS computed with
    and without noise. We have considered a signal-to-noise ratio
    equal to 15, for a CDM simulation without UVB, and with imposed
    temperature in post-processing equal to $T=2\times10^4~{\rm
      K}$. The left (right) panel regards the redshift interval
    centered on $z=5.0$ ($z=5.4$).  The solid lines refers to the mean
    of the ratio between the FPS computed with and without noise, the
    shaded region refer to the $1$-$\sigma$ uncertainty on the ratio.
    The black solid line is the correction for noise applied in
    \citet{Viel13}, that is independent from $\tau_{\rm eff}$. We
    conclude that the effect of noise depends on $\tau_{\rm eff}$, and
    that accounting for noise only with a filter to the noiseless FPS
    is going to introduce a bias in the final estimate of the
    temperature.}
    \label{fig:noise}
\end{figure*}

\section{Estimation of mean flux uncertainties}\label{app:meanfluxerror}

Available measurements of mean flux at high redshifts are based on
small samples of quasars. Data from \citet{Viel13} that we are using
contains only 25 quasars with emission redshifts $4.48 \leq z _ {
  \mathrm { em } } \leq 6.42$. Other works like \cite{Becker:2014oga}
provide mean flux measurements also for only $\sim 10$ redshift
intervals above $z=5$. Even though quoted mean flux errors for
individual spectra can be as low as $\sim 1\%$, tiny sample sizes
suggest that undersampling of the density distribution is occurring.

To estimate this sampling error, we studied the distribution of mean
flux for populations of mock spectra drawn from one of our
simulations. To closely replicate the setup of \cite{Viel13}, from
1000 lines of sight of the length $20\,\text{Mpc}/h$ we prepared 142
l.o.s. of $140\,\text{Mpc}/h$ by random concatenation (roughly
corresponding to $\Delta z = 0.4$ used in \cite{Viel13} to bin the
observations).

Next, we drew 1000 samples of the sizes 1, 10 and 100. For each
population, we computed the standard as well as maximal deviations to
gauge the sampling bias: Table \ref{tab:tau_sampling_errors}. We see
that typical error for $N_{sample}=10$ is of the order of $4-5\%$.

On the other hand, the typical continuum level uncertainty is
estimated to be $\sim 20\%$ \citep{Viel13}. Hence, uncertainty is
dominated by continuum error.

\begin{table}
\centering
\caption{Means and standard deviations for populations of
  $140\,\text{Mpc}/h$ mock spectra. $\bar{F}$ denotes averaging inside
  a population while angular brackets $\left<\right>$ denote ensemble
  average.}
\label{tab:tau_sampling_errors}
\begin{tabular}{llll}
z                    & $N_{\text{sample}}$ & $\left<\bar{F}\right>$          &   Standard Deviation       \\  \hline
\multirow{3}{*}{5.4} & 1                   & $0.1136$ & $\pm0.0163$  $(\pm14.3\%)$  \\
                     & 10                  & $0.1121$ & $\pm0.0056$  $(\pm5.0\%)$  \\
                     & 100                 & $0.1121$ & $\pm0.0008$  $(\pm0.7\%)$   \\  \hline
\multirow{3}{*}{5.0} & 1                   & $0.2086$ & $\pm0.0247$  $(\pm11.8\%)$  \\
                     & 10                  & $0.2070$ & $\pm0.0078$  $(\pm3.8\%)$     \\
                     & 100                 & $0.2065$ & $\pm0.0011$  $(\pm0.5\%)$ 
\end{tabular}
\end{table}
\section{Analysis of the HIRES data-sets at $z=4.2$ and $z=4.6$}\label{app:allz}
In this analysis we have considered the HIRES data-sets at $z=5.0$ and
$z=5.4$ that were already presented in \citep{Viel13}. In that same
work, the authors had analyzed there were two more HIRES data-sets at
$z=4.2$ and $z=4.6$. Because we have previously shown that the
previous constraints on WDM are obtained from the data at $z=5.0$
\citep{Garzilli:2015iwa}, we have focused on analyzing data at $z=5.0$ and
$z=5.4$. Here, for completeness we give the results of our analysis for
these later redshift intervals. In Figure~\ref{fig:otherz} we show the
confidence level for $\tau_{\rm eff}$ and $T_0$ for $z=4.2$ and
$z=5.0$. In Table~\ref{tab:bestfit2} we show the values of the
$\chi^2$ for  the best-fitting models. 

\begin{figure*}
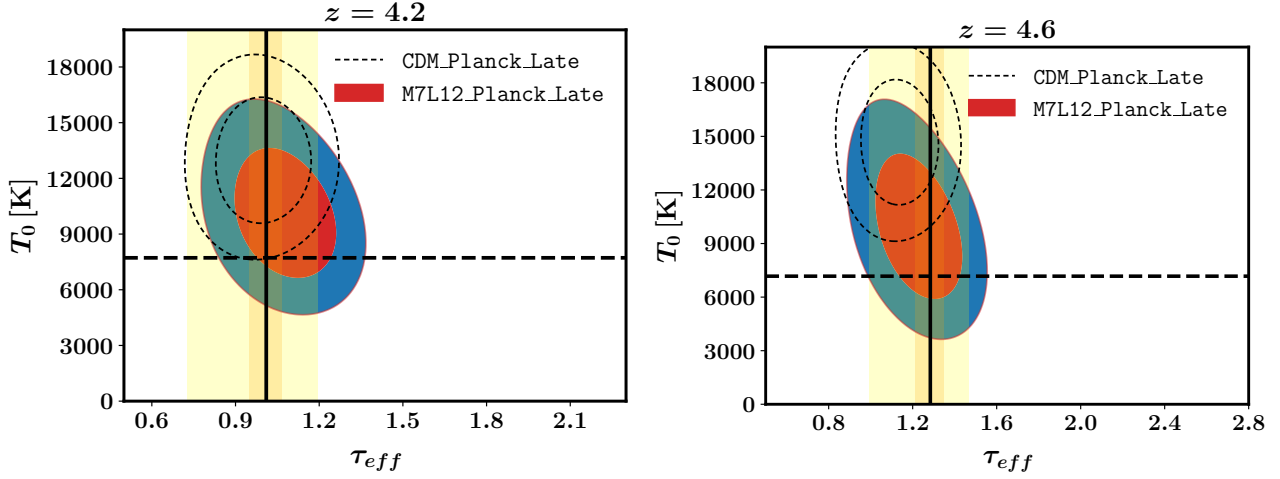

  \includegraphics[width=\columnwidth]{figs/figE1a}
  \includegraphics[width=0.98\columnwidth]{figs/figE1b}
  \caption{Confidence levels of mock FPS compared to the observed FPS
    of {\sc hires} for redshifts $z=4.2$ (left) and $z=4.6$
    (right). Same convention as in Figure~\ref{fig:bichi2}. \label{fig:otherz}
}
\end{figure*}

\begin{table}
\centering
\caption{Values of $\chi^2$ for the  best-fitting models shown in
  Figure~\ref{fig:otherz}. The number of dof is 5.\label{tab:bestfit2}}
\begin{tabular}{l|c|c}
  \hline
  model &$z$ & $\chi^2$ \\
  \hline
  CDM\_Planck\_Late & 4.2 & 9.91 \\ 
      & 4.6 & 4.61 \\
  \hline
  M7L12\_Planck\_Late & 4.2 & 6.04 \\ 
      & 4.6 & 3.99 
\end{tabular}
\end{table}

\bibliography{refs,merged}
\end{document}